\documentclass[11pt,a4paper]{article}
\usepackage{jcappub}
\usepackage{graphicx}	
\usepackage{amsmath, cases}

\title{Blandford\textendash{}Znajek jets in MOdified Gravity}

\author[a,b]{Filippo Camilloni,}
\author[c]{Troels Harmark,}
\author[a,c]{Marta Orselli}
\author[d,e]{\\and Maria J. Rodriguez}

\affiliation[a]{Dipartimento di Fisica e Geologia, Universit\`a di Perugia,
\\
I.N.F.N. Sezione di Perugia,\\
Via Pascoli, I-06123 Perugia, Italy}
\affiliation[b]{Institut f{\"u}r Theoretische Physik, Goethe Universit{\"a}t,
\\
Max-von-Laue-Str. 1, 60438 Frankfurt am Main, Germany}
\affiliation[c]{Niels Bohr International Academy, Niels Bohr Institute, Copenhagen University,\\
Blegdamsvej 17, DK-2100 Copenhagen, Denmark}
\affiliation[d]{Department of Physics, Utah State University,\\
 4415 Old Main Hill Road, UT 84322, U.S.A.}
\affiliation[e]{Instituto de Fisica Teorica UAM/CSIC, Universidad Autonoma de Madrid,\\
13-15 Calle Nicolas Cabrera, 28049 Madrid, Spain}

\emailAdd{camilloni@itp.uni-frankfurt.de}
\emailAdd{harmark@nbi.ku.dk}
\emailAdd{orselli@pg.infn.it}
\emailAdd{maria.rodriguez@usu.edu}

\abstract{General relativity (GR) will be imminently challenged by upcoming experiments in the strong gravity regime, including those testing the energy extraction mechanisms for black holes. Motivated by this, we explore magnetospheric models and black hole jet emissions in Modified Gravity (MOG) scenarios. Specifically, we construct new power emitting magnetospheres in a Kerr-MOG background which are found to depend non-trivially on the MOG deformation parameter. This may allow for high-precision tests of GR.
In addition, a complete set of analytic solutions for vacuum magnetic field configurations around static MOG black holes are explicitly derived, and found to comprise exclusively Heun's polynomials.}

\arxivnumber{2307.06878}

\begin{document}
\maketitle

\section{Introduction}
\label{Sec:introduction}

In recent years new astrophysical observations have given striking confirmations to the prediction of General Relativity (GR). The detection of gravitational waves by the LIGO-Virgo-KAGRA collaboration provided the first direct observations of coalescing binary systems of black holes and neutron stars \cite{LIGOScientific:2016aoc,LIGOScientific:2017vwq,LIGOScientific:2021qlt}, whereas the images revealed by the Event Horizon Telescope (EHT) collaboration confirmed the presence of supermassive black holes harboured in the nuclei of galaxies \cite{EventHorizonTelescope:2019dse,EventHorizonTelescope:2022wkp}, and gave support to the hypothesis that the astrophysical jets associated to Active Galactic Nuclei (AGNs) are powered by spinning black holes via the Blandford--Znajek (BZ) mechanism \cite{Blandford:1977ds,EventHorizonTelescope:2019pgp,EventHorizonTelescope:2021srq}.

The new observational opportunities at our disposal pave the way to the detection of strong-gravity effects in a variety of astrophysical systems,  where potential deviations from GR are expected to manifest. In this regard, gravitational waves detectors and telescope arrays such as the EHT not only constitute a source of new observational discoveries, but also an invaluable ground where to make comparison between precise theoretical predictions of GR and alternative theories of gravity. For instance, the first multimessenger observation of a neutron star merger/GRB event, GW170817/GRB170817A \cite{LIGOScientific:2017zic}, not only confirmed binary neutron star mergers as progenitor of short gamma-ray bursts, but also allowed to impose stringent constraints that ruled out all the theories which are not consistent with the weak form of the equivalence principle \cite{Boran:2017rdn}. 
Similarly, the comparison with EHT observations of the M87* shadow permitted to exclude many models of ``black hole mimickers" as well as black holes with additional massive scalar hairs as candidates for the central object \cite{EventHorizonTelescope:2019pgp}.

Among the candidates that survived observations, the Scalar-Tensor-Vector Gravity (STVG) \cite{Moffat:2005si} -- also referred to as MOdified Gravity (MOG) in the literature -- provides a covariant extension of GR in which electromagnetic and gravitational perturbations propagate with the same speed, thus being consistent with the GW170817/GRB170817A event \cite{Green:2017qcv}, and  admitting spinning black hole solutions \cite{Moffat:2014aja} whose shadows are qualitatively indistinguishable from the one of M87* within the present EHT precision \cite{EventHorizonTelescope:2019pgp}. From a theoretical perspective, the interest for MOG theory stems from the fact that it constitutes a covariant generalisation of GR based on an action principle that offers phenomenological explanations to the galactic rotation curves \cite{Brownstein:2005zz,Moffat:2013sja}, the bullet cluster phenomenon \cite{Brownstein:2007sr} and cluster dynamics \cite{Moffat:2013uaa}, without the necessity of introducing dark matter. In light of this, it is urgent to develop new theoretical signatures that can help to distinguish GR from MOG in present and future astrophysical observations.\\

One of the most characteristic features of spinning black holes is the possibility of extracting rotational energy via the Penrose process \cite{1971NPhS..229..177P,Lasota:2013kia}, whose electromagnetic manifestation, namely the BZ mechanism \cite{Blandford:1977ds}, is considered to be the primal engine behind the emission of relativistic jets in AGN.
Despite its great importance in the field of relativistic astrophysics, the inherently complex dynamics behind the BZ mechanism constituted a major obstacles for theorists to clarify details about its physics, and only in the last two decades progress in this directions were made either by means of numerical simulations \cite{10.1046/j.1365-8711.2001.04863.x,Komissarov:2004ms,McKinney:2004ka,Tchekhovskoy:2011zx,Nathanail:2014aua}, and by means of analytic studies \cite{Tanabe:2008wm,Pan:2015iaa,Grignani:2018ntq,Jacobson:2017xam,Grignani:2019dqc,Armas:2020mio,Camilloni:2020hns,Camilloni:2020qah,Carleo:2022ukm,Camilloni:2022kmx}.

Recently there have been efforts to understand how deviations from GR can manifest in the physics of relativistic jets. Examples range from studies on emission mechanisms in regular black hole metrics \cite{Pei:2016kka}, Kerr-Sen black holes \cite{Banerjee:2020ubc}, quadratic and cubic theories of gravity \cite{Dong:2021yss,Peng:2023}, and on signatures from extra dimensions \cite{Chanson:2022byc}.
A first attempt to investigate how MOG deformations reflect on jet dynamics has been conducted in \cite{LopezArmengol:2016nyi}, where the Blandford $\&$ Payne mechanism \cite{1982MNRAS.199..883B} was assumed to cause the jet launching and particle trajectories around Kerr-MOG black holes were computed.

In this work we consider an alternative path with respect to \cite{LopezArmengol:2016nyi}. We explore the Kerr-MOG scenario by regarding the jet as being powered by the black hole, with the BZ mechanism responsible for the extraction of energy and angular momentum, and by focusing on the dynamics of force-free magnetospheres around MOG black holes.

Indeed, the comparison between EHT observations and General Relativistic MagnetoHydroDynamics (GRMHD) simulations revealed that the power emitted in the launching region of M87* is consistent with the BZ mechanism \cite{EventHorizonTelescope:2019pgp,EventHorizonTelescope:2021srq}.
More specifically, numerical results from GRMHD simulations \cite{2010ApJ...711...50T,Tchekhovskoy:2011zx} showed that the rate of energy extracted in the BZ mechanism from a black hole surrounded by a razor-thin accretion disc can be generally expressed as 
\begin{equation}
\label{eq: BZ_0}
    \dot E^{BZ}=\kappa~\Omega_H^2(2\pi \Psi_H)^2f(\Omega_H)~~,
\end{equation}
where the quantity $(2\pi\Psi_H)$ represents the total flux threading the event horizon, $\Omega_H$ is the black hole angular velocity, and $\kappa$ is a factor taking into account the geometry of the magnetic field configuration ($\kappa=\frac{2\pi}{3}\cdot\frac{1}{4\pi^2}\approx 0.053$ for a monopolar field).
The function $f(\Omega_H)$ is approximately $f(\Omega_H)\approx 1$ for slowly-spinning black holes, reproducing the widely used quadratic scaling for $\dot E^{BZ}$ computed originally in \cite{Blandford:1977ds}, whereas in the high-spin regime its explicit expression is typically given in terms of an expansion in powers of $\Omega_H$. Currently, the most accurate expression for $f(\Omega_H)$, obtained in the case of a monopolar magnetosphere around a Kerr black hole,  was computed analytically up to orders $\mathcal{O}(\Omega_H^6)$ in \cite{Camilloni:2022kmx}, improving previous estimates derived by means of numerical fit with GRMHD simulations \cite{2010ApJ...711...50T} which truncated the series at $\mathcal{O}(\Omega_H^4)$. 
It is crucial to stress that the authors in \cite{2010ApJ...711...50T} gave indications that the function $f(\Omega_H)$ remains the same even for different magnetic field geometries. Being the BZ mechanism dependent on both the configuration of the magnetosphere surrounding a black hole and the theory of gravity considered, one expects that the power extracted from black holes in alternative scenarios can constitute a signature to reveal deviations from standard GR results \cite{Bambi:2012ku,Bambi:2012zg,Pei:2016kka}. In what follows we argue that such deviations are mainly encoded in the function $f(\Omega_H)$.
The main purpose of this work is indeed to derive an explicit expression for $f(\Omega_H)$ in the MOG case beyond the leading order approximation, and show that this function depends in a non-trivial manner on the MOG deformation parameter that accounts for deviations from GR. In doing this we apply the standard BZ perturbation theory to construct analytic models of black hole magnetospheres and to explicitly compute the energy extracted via the BZ mechanism, along the lines of \cite{Armas:2020mio} and \cite{Camilloni:2022kmx}.

More generally, one expects that the expression of $f(\Omega_H)$ is characteristic of the underlying theory of gravity on which the BZ mechanism is set to operate.
Hence, further investigations on the BZ mechanism have the potential to produce new theoretical predictions and observational signatures that would enable current and future horizon-scale measurements done by the EHT collaboration to distinguish GR from alternative theories of gravity, test the Kerr paradigm and clarify details concerning the structure of magnetic fields around spinning black holes.
While future high-precision polarimetric observations with EHT are expected to put more stringent constraints on the magnetic flux and magnetic field geometry threading the black hole \cite{EventHorizonTelescope:2019pgp}, an accurate knowledge of $f(\Omega_H)$ becomes crucial if one aims to use power estimates in order to discriminate between jet models in GR and modified theories of gravity.
As mentioned earlier, in alternative scenarios one might expect that  $f(\Omega_H)$ will depend on one or more additional deformation parameters, considerably increasing the parameter space of the theory. 
Therefore, it is important to pursue an analytical derivation of $f(\Omega_H)$. This is also motivated by the fact that the role of $f(\Omega_H)$ becomes relevant in the high-spin regime where GRMHD simulations are computationally expensive \cite{Talbot:2020zkb}, and analytic results can give important insights for the development of future numerical codes.
\\

The present work is structured as follows. In Sec.~\ref{sec: MOG Kerr} we review some of the main features of STVG and of its Kerr-MOG solution for spinning black holes, whereas in Sec.~\ref{sec: FFE} we collect standard results and equations that captures magnetospheric dynamics in stationary and axisymmetric backgrounds, which find application in the case under study. In Sec.~\ref{sec: static} vacuum magnetic field configurations around static MOG black holes are classified. This will serve as the starting point for the perturbative construction of a spinning monopolar magnetosphere around Kerr-MOG, which we detail in Sec.~\ref{sec: BZMOG}.
An explicit expression for the power and angular momentum extracted in the BZ mechanism, together with a first-order expression for the factor $f(\Omega_H)$ in terms of the MOG deformation parameter is given in Sec.~\ref{sec: BZPower}, where we also make comparisons with the usual BZ theory in GR. We conclude the paper with a discussion in Sec.~\ref{sec: conclusion}.
\\

Where not explicitly specified we adopt geometrised units by setting $G_N = c = 1$ for the Newton's constant and the speed of light, and signature $(-,+,+,+)$ for the metric.

\section{MOdified Gravity (MOG)}
\label{sec: MOG Kerr}

We provide in this section a short review of the STVG theory, aimed at clarifying its main features and the motivation behind its construction. We first list the various fields characterising this theory and write the action explicitly. 
We then focus on presenting the Kerr-MOG black hole, a  stationary and axisymmetric solution of the STVG equations of motion which, for the rest of the paper, will be regarded as the background to construct magnetospheric models and study the BZ mechanism.

\subsection{Scalar-Tensor-Vector Gravity action}

STVG has been developed with the aim of providing a fully relativistic generalisation of GR, in which the weak-field modified gravity effects become appreciable over galactic scales, thus explaining astrophysical phenomena, as for example the galaxy rotation curves, without postulating the existence of non-baryonic dark matter.

The main idea of STVG consists in enhancing the gravitational coupling constant $G$ to a dynamical scalar field whose asymptotic value exceeds the Newton constant, $G_N=6,67\times 10^{-11}~{\rm Nm^2/kg^2}$. In order to compensate for the increased value of $G$ at scales comparable to the solar system, where GR provides an accurate description of the physics we observe, a \emph{repulsive} Yukawa interaction is included in the theory. This is achieved by adding a Proca massive vector field $\phi_\mu$ to the action, so that, at  short scales, the value of $G$ reduces to $G_N$ and STVG effectively reconciles with GR and its Newtonian limit without violating any known local observations.

One of the main advantages of MOG theory is that it descends from an action principle. The action consists of four contributions \cite{Moffat:2005si,Moffat:2013sja,Moffat:2013uaa}
\begin{equation}
    S=S_{g}+S_\phi+S_s+ S_M~~,
    \label{actMOG}
\end{equation}
where 
\begin{equation}
S_{g}=\frac{1}{16\pi}\int \frac{1}{G}R~\sqrt{-g}d^4x~~,
\end{equation}
with $R$ being the Ricci scalar computed from the dynamical metric $g_{\mu\nu}$. The additional contributions $S_\phi$ and $S_s$ are respectively associated to the vector and scalar fields of the theory, explicitly given by \footnote{For simplicity we do not consider the presence of the cosmological constant, and set $\omega(x)=1$ as well as $V(\omega)=0$ for the dimensionless scalar field present in the original version of the STVG action presented in \cite{Moffat:2005si}. This choice is consistent and employed, for instance, in \cite{Moffat:2013uaa} and \cite{Moffat:2014aja}.}
\begin{equation}
\label{eq: S_princ}
    \begin{split}
        S_\phi&=-\int \left(\frac{1}{4}B^{\mu\nu}B_{\mu\nu}-\frac{\mu^2}{2}\phi^\mu\phi_\mu+V(\phi)\right)~\sqrt{-g}d^4x~~,
        \\
        S_s&=\int\frac{1}{G}\left[g^{\mu\nu}\left(\frac{\partial_\mu G\partial_\nu G}{G^2}+\frac{\partial_\mu \mu\partial_\nu \mu}{\mu^2}\right)-\left(\frac{V(G)}{G^2}+\frac{V(\mu)}{\mu^2}\right)\right]~\sqrt{-g}d^4x~~ .
    \end{split}
\end{equation}
In the expressions above the quantity $B_{\mu\nu}\equiv2\partial_{[\mu}\phi_{\nu]}$ represents the field strength for $\phi^\mu(x)$, coupled to the dynamical scalar field $\mu(x)$, whose vacuum state plays the role of an effective mass for the vector field, and set the scale at which the deviations from GR become appreciable. As mentioned earlier, in STVG the effective gravitational coupling is promoted to a dynamical scalar field $G(x)$, whose value is in general allowed to vary in space and time. The quantities $V(\phi),~V(G)$ and $V(\mu)$ denote possible self-interactions of the fields. 
Finally the term $S_M$ represents the action contribution associated to source matter fields, which are assumed to couple with both the metric $g_{ \mu\nu}$ and the vector field $\phi^\mu$.

\subsection{Kerr-MOG black holes}

A vacuum static and spherically symmetric black hole solution for STVG has been found in \cite{Moffat:2005si,Moffat:2007nj} by solving the field equations associated to the action \eqref{actMOG}. The solution was later generalised to the stationary and axisymmetric case in \cite{Moffat:2014aja}, and denoted as the \emph{Kerr-MOG black hole}. In obtaining such solutions one neglects self-interaction potentials by setting $V(\phi)=V(G)=V(\mu)=0$, and fix constant values for the scalar fields $G$ and $\mu$. In particular, one assumes that the asymptotic value of the enhanced gravitational constant, $G\equiv G_\infty=(1+\alpha)G_N$, differs from the Newtonian constant $G_N$ for a deformation parameter
\begin{equation}
\label{eq: MOG_alpha}
    \alpha=\frac{G-G_N}{G_N}~~.
\end{equation}
The asymptotic solution for the vector field $\phi^\mu$ reveals that its time component is a Yukawa-type potential \cite{Moffat:2005si}
\begin{equation}
\label{eq: yukawa}
    \phi_t^\infty= -K \frac{e^{-\mu r}}{r}~~,
\end{equation}
whose charge $K$ is \emph{postulated} \cite{Moffat:2014aja} to acquire a gravitational character by means of the relation
\begin{equation}
    K=\sqrt{\alpha G_N} M~~,
\end{equation}
in order to recover the Newtonian value for $G$ in the proximity of a source mass $M$. Assuming a minimal coupling between test particles and the vector field, it was shown that the repulsive Yukawa term \eqref{eq: yukawa} adds to the Newtonian gravitational potential, leading to a modified Newtonian dynamics in the weak field regime of the MOG background \cite{Moffat:2005si,Moffat:2013sja}.\\
The mass value $\mu$ can be fixed in such a way that effects of deviations from GR -- namely the cut-off scale for the Yukawa potential -- manifest at distances of kiloparsecs from the gravitational source. The value $\mu\approx2.6\times 10^{-28}~ {\rm eV}$, corresponding to a lengthscale $\ell=\mu^{-1}\approx 23.8~ {\rm kpc}$, proved to correctly reproduce the galaxy rotation curves
\cite{Brownstein:2005zz,Moffat:2013sja}, the bullet cluster \cite{Brownstein:2007sr} and cluster dynamics
\cite{Moffat:2013uaa}, but can be effectively neglected in the derivation of black hole solutions \cite{Moffat:2014aja}. Under this approximation the vector field leads to an additional repulsive Coulomb contribution to the gravitational potential.\\

The Kerr-MOG spacetime \cite{Moffat:2014aja} is a stationary and axisymmetric three-parameters family of solution for STVG, which describes a black hole characterised by an angular momentum and an additional gravitational charge reflecting the presence of the additional vector field. The spacetime metric $g_{\mu\nu}$ and the vector field $\phi^\mu$ can be expressed in Boyer-Lindquist coordinates $x^\mu=(t,r,\theta,\varphi)$ in terms of the line element $ds^2=g_{\mu\nu}dx^\mu dx^\nu$, namely
\begin{equation}
\label{eq:KerrMOG}
    \begin{split}
        ds^2&=-\frac{\Delta \Sigma}{\Xi}dt^2+\frac{\Sigma}{\Delta}dr^2+\Sigma d\theta^2+\frac{\Xi \sin^2 \theta}{\Sigma}(d\varphi-\omega dt)^2~~,
        \\
        \phi_\mu&=\frac{K r}{\Sigma}\left(dt-a \sin^2\theta~ d\varphi\right)_\mu~~,
    \end{split}
\end{equation}
where, adopting units for which $G_N=1$, one has
\begin{equation}
\begin{split}
    \Sigma&=r^2+a^2\cos^2\theta~,
    ~~
    \qquad \Delta=r^2-2M_{\alpha}  r+a^2+ (1+\alpha) K^2~,
    \\
    \omega &=a\frac{(2M_{\alpha}r-K^2)}{\Xi}~,
    ~~
    \quad\Xi=(r^2+a^2)^2-a^2\Delta \sin^2\theta~,
\end{split}
\end{equation}
with $a$ being the angular momentum per unit mass. Accordingly, the Arnowitt-Deser-Misner (ADM) \cite{PhysRev.116.1322} mass, angular momentum and gravitational charge for the Kerr-MOG black hole \cite{Sheoran:2017dwb} are defined as 
\begin{equation}
\label{eq: ADM}
    M_{\alpha}=(1+\alpha)M~,~~J=a M_\alpha~,~~K=\frac{\sqrt{\alpha }}{1+\alpha}M_\alpha~~.
\end{equation}
It is important to remark that, while the parameter $M$ only serves as an integration constant of the STVG field equations, it is $M_\alpha$ that corresponds to the gravitational mass of the Kerr-MOG black hole, in accordance with the ADM Hamiltonian formulation \cite{Sheoran:2017dwb}. We notice that the presence of a non-vanishing MOG deformation parameter $\alpha$, defined in Eq.~\eqref{eq: MOG_alpha}, leads to a difference between $M$ and the ADM mass $M_\alpha$.

In the following it will be convenient to introduce a dimensionless spin parameter, given by 
\begin{equation}
    \epsilon=\frac{a}{M_\alpha}~~,
\end{equation}
and for the rest of the paper we assume $a$, or analogously $\epsilon$, to be non-negative. In the limit $\epsilon\to0$ the metric \eqref{eq:KerrMOG} describes a static MOG black hole \cite{Moffat:2005si,Moffat:2014aja}, whereas for $\alpha\to0$ it smoothly reduces to the Kerr solution of standard GR.

The Kerr-MOG metric \eqref{eq:KerrMOG} admits an inner and an outer horizon, denoted here as $r_{\pm}$ respectively, solutions of $\Delta=0$, and respectively located at
\begin{equation}
    \label{eq:EventHorizon}
    r_{\pm}=M_{\alpha}\left(1\pm\sqrt{\frac{1}{1+\alpha}-\epsilon^2}\right).
\end{equation}
It is possible to associate an angular velocity to the event horizon $r_+$, given by 
\begin{equation}
    \label{eq:Omega_H}
    \Omega_H=\omega(r_+,\theta)=\frac{\epsilon}{ 2r_+-\alpha(1+\alpha)^{-1}M_\alpha}~~.
\end{equation}
The boundary of the ergoregion, or ergosurface, can simply be defined as the locus of spacetime points where the asymptotic timelike Killing field $\partial_t$ becomes null, \emph{i.e.} where $g_{tt}=0$. Consequently, the ergosurface $r_{\rm e}(\theta)$ is found at
\begin{equation}
    \label{eq:ergosurf}
    r_{\rm e }(\theta)=M_{\alpha}\left(1+\sqrt{\frac{1}{1+\alpha}-\epsilon^2\cos^2\theta}\right).
\end{equation}
By requiring the gravitational charge $K$ in Eq.~\eqref{eq: ADM} and the event horizon position $r_+$ as defined in Eq.~\eqref{eq:EventHorizon} to be real quantities, in accordance with the cosmic censorship conjecture, one obtains physical bounds on the MOG deformation parameter $\alpha$~\cite{Sheoran:2017dwb,Guo:2018kis}
\begin{equation}
\label{eq:PhysicalBound}
0\leq \alpha\leq\epsilon^{-2}-1~~.
\end{equation}
In particular extremality, is defined when the equality holds in the second relation above, meaning that the two horizons collapse in a single spherical surface of radius $r_\pm=M_\alpha=|a|\sqrt{1+\alpha}$. Hence, a non-vanishing value for the deformation parameter $\alpha$ implies that, for the same ADM mass and angular momentum, the size of the horizon for a Kerr black hole is always bigger than the one of its MOG counterpart.
Notice also that, unlike in the Kerr metric, having that the specific angular momentum equates the ADM mass is not possible in Kerr-MOG. In order to make a comparison with the situation in GR, we label respectively with $M_\star$ and $J_\star$ the ADM mass and angular momentum of a GR Kerr black hole. Equating the ADM mass with its Kerr-MOG counterpart means $M_\star=M_\alpha$ and it is immediate to notice that, while for extreme Kerr $J_\star = M_\star^2$, one has $J = M_\alpha^2/\sqrt{1+\alpha} < J_\star$ in Kerr-MOG. Thus, by spinning up a Kerr and a Kerr-MOG black holes of same ADM masses one would reach extremality in the MOG scenario before than in the GR case.

\section{Black hole magnetospheres dynamics}
\label{sec: FFE}

In this section we review conventions and equations that will be pivotal for extending the BZ model to the case of Kerr-MOG black holes.
In particular, we assume that a vacuum-breaking process similar to that outlined by Blandford and Znajek \cite{Blandford:1977ds} exists so that at equilibrium a plasma-filled magnetosphere, represented by a probe electromagnetic field $F_{\mu\nu}$ sustained by a current density of the plasma $j^\mu$, surrounding the Kerr-MOG black hole establishes.
As a first exploration of the magnetospheric problem in a MOG background, we also assume that an eventual coupling between electromagnetic fields and the MOG vector field $\phi_\mu$ is negligible~\footnote{This assumption can be in principle relaxed, as shown for instance in \cite{Rahvar:2018nhx}, where the authors \emph{postulate} a coupling between the electromagnetic stress-energy tensor and the MOG vector field, leading to modified equations for Maxwell's electrodynamics. }. This is to avoid that the Kerr-MOG background can act as a source for additional electromagnetic fields, and guarantees that Maxwell's equations are still given by $\nabla_{[\rho}F_{\mu\nu]}=0$ and $\nabla_{\nu}F^{\mu\nu}=j^{\mu}$, with $\nabla_\mu$ being the covariant derivative defined with respect to the Kerr-MOG metric in Eq.~\eqref{eq:KerrMOG}. 

\subsection{Force-Free Electrodynamics (FFE)}

The minimum non-trivial level of description for magnetospheres around compact objects like black holes and neutron stars is correctly captured by the theory of \emph{Force-Free Electrodynamics} (FFE) \cite{Blandford:1977ds,PhysRevE.56.2181, Gralla:2014yja}, a regime of magnetohydrodynamics in which most of the energy resides in the electromagnetic sector of the system and the inertia, as well as the thermal degrees of freedom, of the plasma can be effectively neglected \cite{1997PhyU...40..659B}. Under these assumptions the magnetic fields dominates the energy and momentum balance of the system, and the stress-energy tensor respects the \emph{force-free approximation} $T_{\mu\nu}\simeq T^{EM}_{\mu\nu}$.

The FFE system of equations is therefore given by the Maxwell's equations $\nabla_{[\rho}F_{\mu\nu]}=0$, and $\nabla_{\nu}F^{\mu\nu}=j^{\mu}$ supplemented with the \emph{force-free condition}
\begin{equation}
    F_{\mu\nu}j^{\nu}=0~,~~j^\nu\neq0~~.
\end{equation}
By means of the non-homogeneous Maxwell's equations, and assuming the field to be magnetically-dominated, $F^{\mu\nu}F_{\mu\nu}>0$ \cite{PhysRevE.56.2181,Gralla:2014yja, Komissarov:2004ms}, it is possible to formally eliminate the current $j^\mu$ from the dynamical description and summarise the FFE equations as
\begin{equation}
\label{eq:FFE}
    F_{\mu\nu}~\nabla_{\rho}F^{\nu\rho}=0~,~~\nabla_{[\rho}F_{\mu\nu]}=0~~.
\end{equation}
Accordingly, FFE is a non-linear regime of Maxwell's electrodynamics which is only specified by the electromagnetic field configurations, whereas the current $j^\mu$ can be regarded as a secondary quantity that descends from the fields, which only serves to sustain the magnetic fields and screen the longitudinal component of the electric fields.
\\

We consider magnetospheres around a Kerr-MOG black hole that share the same symmetries of the background, \emph{i.e.} that are stationary and axisymmetric \cite{Blandford:1977ds,PhysRevE.56.2198}. 
As usual in magnetohydrodynamics, stationary and axisymmetric flows are characterised by a set of integrals of motion taking constant values along the magnetic field lines \cite{Beskin:2004zc}. More specifically, the \emph{magnetic flux} across a circular loop of radius $r\sin\theta$ piercing the rotational axis of the black hole, $\Psi=\Psi(r,\theta)$, is constant along the projection of the field lines on the poloidal plane $(r,\theta)$. In the force-free limit the electromagnetic field is totally determined by two additional integrals of motion \cite{Gralla:2014yja},  respectively representing the \emph{poloidal current} $I=I(r,\theta)$ flowing through a loop around the rotational axis, and $\Omega=\Omega(r,\theta)$, namely the \emph{angular velocity of the magnetic field lines}. These two are subject to integrability conditions
\begin{equation}
    \label{eq: intcond}
    \partial_r I \partial_\theta \Psi=\partial_\theta I \partial_r \Psi~~,~~\partial_r \Omega \partial_\theta \Psi=\partial_\theta \Omega \partial_r \Psi~~,
\end{equation}
which descend from Eq.~\eqref{eq:FFE}, and imply that $I\equiv I(\Psi)$ and $\Omega\equiv \Omega(\Psi)$ \cite{Blandford:1977ds,1997PhyU...40..659B,Camilloni:2022kmx}.\\
As anticipated, a generic stationary and axisymmetric force-free field in the Kerr-MOG background is written solely in terms of the integrals of motion, and reads \cite{Gralla:2014yja}
\begin{equation}
    \label{eq: F_FFE}
    F=d\Psi\wedge \eta-I(\Psi)~\frac{\Sigma}{\Delta\sin\theta}\,dr\wedge d\theta~~,
\end{equation}
where the one-form 
\begin{equation}
    \label{eq:eta}
    \eta=d\varphi-\Omega(\Psi)~dt~~,
\end{equation}
is defined to be always orthogonal to the four-velocity of a particle which is co-rotating together with the field lines, with an angular velocity $\Omega(\Psi)$.
\\

The magnetospheric problem amounts to find explicit functional expressions for the field variables $\Psi(r,\theta)$, $I(\Psi)$ and $\Omega(\Psi)$ on a given background.

\subsection{Grad-Shafranov equation and critical surfaces}

Due to the structural analogy between Kerr and Kerr-MOG it is possible to obtain similar relations that relate the force-free field variables in the stationary axisymmetric case. In this section, therefore, we limit to record the standard set of relations used in the construction of magnetospheric models. For more detailed derivations and discussions we refer the reader to previous work in the literature \cite{Blandford:1977ds,Beskin:2004zc,Gralla:2014yja,Camilloni:2022kmx}.\\

As typical for stationary and axisymmetric flows in magnetohydrodynamics, the field variables of the magnetospheric problem are put in relation altogether by means of the \emph{Grad-Shafranov equation} \cite{1997PhyU...40..659B,Beskin:2004zc}. In the Kerr-MOG spacetime this can be compactly written in a semi-covariant manner \cite{Camilloni:2022kmx} as
\begin{equation}
    \label{eq: GS_eq}
    \eta_\mu \partial_r \Big( \eta^\mu  \Delta \sin \theta \,  \partial_r \Psi \Big) + \eta_\mu \partial_\theta \Big( \eta^\mu  \sin \theta \,  \partial_\theta \Psi \Big) + \frac{\Sigma}{\Delta \sin \theta} I \frac{d I}{d\Psi} =0~~.
\end{equation}
The equation above, also called \emph{stream equation}, emerges by combining the $r$ and $\theta$ components of the force-free condition, Eq.~\eqref{eq:FFE}, in the Kerr-MOG background, together with the field structure \eqref{eq: F_FFE} \cite{Camilloni:2022kmx}.
Besides specifying boundary conditions, which allow to select a particular solution, the Grad-Shafranov equation must also be supplemented with regularity conditions at its critical surfaces \cite{Uzdensky:2003cy,Uzdensky:2004qu,Nathanail:2014aua}. Similar to the case of the standard Kerr spacetime in Boyer-Lindquist coordinates, for Kerr-MOG black holes the magnetospheric structure is featured with the presence of four critical surfaces located at the event horizon, at asymptotic infinity and at two light surfaces.\\

In order to demand regularity of the electromagnetic field at the horizon and at  infinity, the following two \emph{Znajek conditions} \cite{Komissarov:2004ms,Nathanail:2014aua,Armas:2020mio,Camilloni:2022kmx} must be imposed
\begin{equation}
    \label{eq: ZC}
\begin{split}
    I_+(\theta)&=\frac{2M_\alpha r_+}{\Sigma_+}\sin\theta\left(\Omega_H-\Omega_+\right)\partial_\theta\Psi_+~~,
    \\
    I^\infty(\theta)&=\sin\theta\,\Omega^\infty(\partial_\theta\Psi)^\infty~~,
\end{split}
\end{equation}
where the labels $+,\infty$ have been respectively used to evaluate quantities as the radial coordinate approaches the horizon and  infinity ( \emph{e.g.} $I_+\equiv I\vert_{r=r_+}$ and $I^\infty\equiv I\vert_{r=\infty}$), and with $\Omega_H$ being the black hole angular velocity, as given in Eq.~\eqref{eq:Omega_H}.

The light surfaces are defined as the locus of points where the velocity of an observer co-rotating with the field lines becomes null, thus satisfying the condition $g^{\mu\nu}\eta_\mu\eta_\nu=0$ \cite{Gralla:2014yja}. Such an equation generally admits two real and distinct solutions which respectively define the \emph{Inner/Outer Light Surface} (ILS/OLS). At both light surfaces the following Robin-type condition \cite{Uzdensky:2003cy,Uzdensky:2004qu,Gralla:2014yja,Camilloni:2022kmx} 
\begin{equation}
    \label{eq: stream_LS}
    \Delta\,   \eta_\mu \partial_r \eta^\mu  \partial_r \Psi + \eta_\mu \partial_\theta  \eta^\mu  \partial_\theta \Psi  + \frac{\Sigma}{\Delta \sin^2 \theta} I \frac{dI}{d\Psi} =0~~,
\end{equation}
dubbed \emph{reduced stream equation}, must hold. This ensures the smoothness of the magnetic flux $\Psi$ when the magnetic field lines cross these critical surfaces. Detailed information about the properties of the ILS and OLS in the case of the Kerr background can be found in \cite{Komissarov:2004ms}, and they directly extend also to the Kerr-MOG case due to the structural analogy between the two metrics. While the ILS is a closed surface that is always comprised between the event horizon and the ergosphere, and therefore is a distinguishing feature of black hole magnetospheres, the OLS is characterised by an open topology, that constitutes the black hole analogue of pulsar's light cylinder \cite{Nathanail:2014aua}.
\\

We conclude this section by stressing that, as in the GR case, Eq.s~\eqref{eq: GS_eq}, \eqref{eq: ZC} and \eqref{eq: stream_LS}, supplemented with appropriate boundary conditions which specify a particular topology for the magnetic field configuration, constitute the complete set of equations needed to address the magnetospheric problem. The authors of 
\cite{Armas:2020mio} first showed the crucial role of the light surfaces in the construction of a consistent solution for a monopolar magnetosphere around a Kerr black hole by exploiting perturbative techniques enhanced with a matched asymptotic expansion scheme. In \cite{Camilloni:2022kmx} such a methodology has been improved and used to derive semi-analytical results leading to the computation of new perturbative contributions in the power extracted by means of the BZ mechanism, that solved previous discrepancies between numerical simulations and analytic approaches. The results of \cite{Camilloni:2022kmx} are pivotal for a possible understanding of the non-perturbative structure of the BZ theory.
For a comprehensive discussion on the Grad-Shafranov equation and the regularity conditions at the critical surfaces in the context of Kerr black holes we refer the reader to \cite{Camilloni:2022kmx}.

\section{Vacuum solutions in static MOG backgrounds}
\label{sec: static}

The BZ perturbative approach \cite{Blandford:1977ds,Armas:2020mio,Camilloni:2022kmx} consists in building perturbations around vacuum solutions in static black hole backgrounds. It is therefore essential to derive and classify the vacuum electromagnetic field configurations surrounding a static MOG black hole, so as to use them as a starting point to extend the BZ perturbation theory to the case of a Kerr-MOG background. In this section we therefore focus on static MOG black holes which we can describe by setting the spin parameter $\epsilon=0$ in  Eq.~\eqref{eq:KerrMOG}. The metric, thus, reads 
\begin{equation}
\label{eq: SchwMOG}
    ds^2=-\frac{\bar\Delta(r)}{r^2}dt^2+\frac{r^2}{\bar\Delta(r)}dr^2+r^2(d\theta^2+\sin^2\theta d\varphi^2)
\end{equation}
with $\bar\Delta=(r-\bar r_+)(r-\bar r_-)$ and the two horizons in the static case are located at $\bar r_\pm=M_\alpha(1\pm\sqrt{(1+\alpha)^{-1}})$. It worths noticing that, as a consequence of the fact that the ``charge" appearing in the Kerr-MOG metric is tied to the gravitational mass, no extreme limit for a static black hole exists in MOG, as opposed to the Reissner-Nordstr\"{o}m solution in GR.\\

Vacuum solutions, characterised by $j^\mu=0$, can be constructed by assuming $I(\Psi)=\Omega(\Psi)=0$ and by demanding the magnetic flux $\Psi$ to obey the Grad-Shafranov equation \eqref{eq: GS_eq} in the static MOG background. The latter reduces to the homogeneous equation $\mathcal{L}\Psi(r,\theta)=0$, where the Laplace operator in the static MOG background reads
\begin{equation}
    \mathcal{L}=\frac{1}{\sin\theta}\partial_r\left(\frac{\bar\Delta}{r^2}~\partial_r\right)+\frac{1}{r^2}\partial_\theta\left(\frac{1}{\sin\theta}~\partial_\theta\right)~~.
\end{equation}
By assuming a solution of the form $\Psi(r,\theta)\sim R^{(\ell)}(r)\Theta_{\ell}(\theta)$ it is possible to separate the variables, with the radial harmonics $R^{(\ell)}(r)$ and the angular harmonics $\Theta_\ell(\theta)$ being eigenfunctions of two independent Sturm-Liouville problems \cite{Gralla:2015vta}, respectively defined by 
\begin{equation}
    \label{eq: SL_prob}
    \begin{split}
        \mathcal{L}_r^{(\ell)}\big[R^{(\ell)}\big]&=\frac{d}{dr}\left(\frac{\bar\Delta}{r^2}~\frac{dR}{dr}^{(\ell)}\right)-\frac{\ell(\ell+1)}{r^2}R^{(\ell)}=0~~,
        \\
        \mathcal{L}_\theta^{(\ell)}\big[\Theta_\ell\big]&=\frac{d}{d\theta}\left(\frac{1}{\sin\theta}~\frac{d\Theta_\ell}{d\theta}\right)+\frac{\ell(\ell+1)}{\sin\theta}\Theta_\ell=0~~.
    \end{split}
\end{equation}

In the rest of the paper we consider \emph{split-field configurations} \cite{Gralla:2014yja}, and only focus on the domain $\theta=[0,\pi/2]$. Solutions in the entire space can directly be obtained by reflection across the equator $\theta=\pi/2$. In what follows we obtain analytic expressions for the solutions of the two Sturm-Liouville problems \eqref{eq: SL_prob}; in particular, the magnetic flux solution of the vacuum equation is constructed by combining the radial and angular harmonics, according to 
\begin{equation}
    \Psi(r,\theta)=c_0+R^{(\ell)}(r)\Theta_{\ell}(\theta)~~,
\end{equation}
where an integration constant $c_0$ is included and the summation on the index $\ell$ is implicit. The regular solutions we consider are subject to the following boundary conditions \cite{Nathanail:2014aua,Armas:2020mio,Camilloni:2022kmx}
\begin{equation}
\label{eq: BC0}
    \Psi\vert_{\theta=0}=0~,~~\partial_\theta\Psi\vert_{\theta=0}=0~,~~\Psi\vert_{\theta=\frac{\pi}{2}}=1~~,
\end{equation}
with the first two conditions demanding regularity of the field at $\theta=0$, and the last being a normalisation condition on the total magnetic flux passing through a surface encompassing the event horizon.
Later in this section we show that an additional boundary condition in the asymptotic region is needed in order to specify the topology of the vacuum magnetic field \cite{Grignani:2019dqc}.

\subsection{Angular eigenfunctions}

Eigenfunctions for the angular part of the Laplace operator \eqref{eq: SL_prob} have been extensively studied in the literature \cite{Gralla:2015vta}, and here we limit to summarise their main properties. Angular harmonics obeying $\mathcal{L}_\theta^{(\ell)}[\Theta_\ell]=0$ where $\ell$ is a positive integer, 
reduce to a degenerate case of the hypergeometric functions \cite{abramowitz+stegun}. 
One of the solutions is characterized by its irregularity at $\theta=0$. On the other hand, the second solution takes the form of a truncated polynomial that terminates after a finite number of terms. These two types of solutions represent distinct mathematical representations of the behavior of the system. The choice between these solutions will be based on the regularity condition at $\theta=0$, Eq.~\eqref{eq: BC0}.

The regular solution $\Theta_\ell$ can be directly expressed in terms of Gegenbauer polynomials $C_{\ell-1}^{(3/2)} (x)$ \cite{Ghosh:1999in,Gralla:2015vta}. More specifically
\begin{numcases}
	{\Theta_\ell(\theta)=}
		-\frac{\Gamma\left(-\frac{\ell}{2}\right)\Gamma\left(\frac{\ell+1}{2}\right)}{2\sqrt{\pi}}
		\sin^2{\theta}\, C^{(3/2)}_{\ell-1}(\cos{\theta})
		& $\ell$ odd,\nonumber\\
		&\\
		-(-1)^{\ell/2}\frac{\sqrt{\pi}\,\Gamma\left(\frac{\ell}{2}\right)}{4\,\Gamma\left(\frac{\ell+3}{2}\right)}
		\sin^2{\theta}\,C^{(3/2)}_{\ell-1}(\cos{\theta})
		&$\ell$ even.\nonumber
\end{numcases}
where the $C_\ell^\alpha (x)$ are orthogonal polynomials in the interval $x\in [-1,1]$ with respect to the weight function $(1-x^2)^{\alpha - 1/2}$, with $\alpha>1/2$, namely
\begin{equation}
\label{ortogonality}
    \int_0^\pi d\theta\, C_\ell^{(3/2)}(\cos\theta) C_{\ell'}^{(3/2)}(\cos\theta) \sin^3\theta=\frac{(\ell+2)(\ell+1)}{\ell+3/2}\delta_{\ell\ell'}.
\end{equation}
It can be noticed that for all values of $\ell$, one finds a regular solution satisfying $\Theta_\ell\vert_{\theta=0}=0$ and $\partial_\theta\Theta_\ell\vert_{\theta=0}=0$. Instead, at the equator $\theta=\pi/2$, one has that $\Theta_{2\ell}\vert_{\theta=\pi/2}=0$ and $\Theta_{2\ell+1}\vert_{\theta=\pi/2}=1$. 
\\
Below we list the explicit expression for $\ell=0,1,2,3$, which are the only relevant for the present discussion
\begin{equation}
\begin{split}
    &\Theta_0(\theta)=\cos\theta~~,~~
    \Theta_1(\theta)=\sin^2\theta~~,
    \\
    &\Theta_2(\theta)=\cos\theta \sin^2\theta~~,~~
    \Theta_3(\theta)=(1-5\cos^2\theta)\sin^2\theta~~.
\end{split}
\end{equation}

The first solution, in particular, is related to the monopole field configuration that we will mostly concentrate on in this paper.

\subsection{Radial eigenfunctions}
\label{sec: Heun}

In order to study the radial part of the Laplace operator, it is convenient to redefine $\mathcal{L}_r^{(\ell)}$ in terms of a dimensionless radial coordinate 
\begin{equation}
\label{eq: w_def}
    w=\frac{r}{\bar r_+}=\frac{\sqrt{1+\alpha}}{1+\sqrt{1+\alpha}}~\frac{r}{M_\alpha}~~,~~w_a=\frac{\bar r_-}{\bar r_+}=\frac{1+\alpha-\sqrt{1+\alpha}}{1+\alpha+\sqrt{1+\alpha}}~~.
\end{equation}
This shows that the radial equation $\mathcal{L}_r^{(\ell)}[R^{(\ell)}]=0$ is actually a second-order linear Ordinary Differential Equation (ODE) of the kind 
\begin{equation}
    \label{eq: ODE_r}
    R''(w)+p(w)R'(w)+q(w)R(w)=0~~,
\end{equation}
where $\bar \Delta(w)=\bar r_+^2(w-1)(w-w_a)$ with $w_a<1$, and
\begin{equation}
    \begin{split}
        p(w)&=\frac{\bar \Delta'(w)}{\bar \Delta(w)}-\frac{2}{w}=\frac{1}{w-1}+\frac{1}{w-w_a}-\frac{2}{w}~~,
        \\
        q(w)&=-\frac{\bar r_+^2\ell(\ell+1)}{\bar\Delta(w)}=-\frac{\ell(\ell+1)}{(w-1)(w-w_a)}~~.
    \end{split}
\end{equation}
This makes manifest the fact that Eq.~\eqref{eq: ODE_r} is in general characterised by four regular singular points, located at $w=0,w_a,1,\infty$, with $1> w_a\geq0$ due to the coordinate choice adopted. It is immediate to notice that the Laplacian operator in the standard GR case ($\alpha\to0$) is recovered by simply taking the limit $\bar r _-\to0$ and $\bar r _+\to2 M$, in such a way that $w_a\to0$.\\

Every linear second-order ODE with at most 4 regular singularities in the complex plane can always be reduced to the \emph{Heun's Equation} \cite{Ronveaux:1995:HDE}
\begin{equation}
    \label{eq: Heuneq}
    R''(w)+\left(\frac{\gamma}{w}+\frac{\delta}{w-1}+\frac{\varepsilon}{w-w_a}\right)R'(w)+\frac{\lambda \chi ~w -q}{w(w-1)(w-w_a)}R(w)=0~~,
\end{equation}
where the complex number $q$ is called the accessory parameter, and with each of the 4 regular singularities $w=0,1,w_a,\infty$ related to a pair of characteristic exponents, respectively given by $(0,1-\gamma)$, $(0,1-\delta)$, $(0,1-\varepsilon)$ and $(\lambda, \chi)$. In order to guarantee that  infinity constitutes a regular singular point, the parameter $\varepsilon$ is subject to the Fuchsian constraint \cite{Ronveaux:1995:HDE}
\begin{equation}
    \varepsilon=1+\lambda+\chi -\gamma-\delta~~,
\end{equation}
so that the Heun's equation is specified by the set of 6 parameters $w_a,q,\delta,\gamma,\lambda,\chi$. We notice that the Heun's equation and its confluent versions have often found applications in  black hole physics (see Ref.s~\cite{Batic:2007it,Fiziev:2009wn,Castro:2013lba,Lupsasca:2014pfa,Lupsasca:2014hua,Compere:2015pja} for a partial literature).

In the case considered here, Eq.~\eqref{eq: ODE_r} directly reduces to the Heun's equation \eqref{eq: Heuneq} upon identifying
\begin{equation}
    \gamma=-2~,~~\delta=1~,~~\varepsilon=1~,~~\lambda=-(1+\ell)~,~~\chi=\ell~,~~q=0~~.
\end{equation}
As mentioned above, in the limit $w_a\to0$ the standard Schwarzschild case in GR can be recovered and, accordingly, the Heun's equation reduces to the Papperitz-Riemann equation, whose solutions are given in terms of hypergeometric functions \cite{Gralla:2015vta}.
\\

Similarly to what one does in the standard Schwarzschild case, we look for two families of polynomials, which arise as Frobenius solutions of the Heun's equation \eqref{eq: Heuneq}.
There are $2\times4\times6$ Frobenius solutions at each singular point, for a total of $192$ Frobenius' solutions \cite{2007MaCom..76..811M}. The symbol $H_l(w_a,q,\lambda,\chi,\delta,\gamma; w)$ is typically used to label the solution at $w=0$ with characteristic exponent 0. All the remaining solutions can be determined by acting with M\"obius and indeces transformations on the Heun's equation \eqref{eq: Heuneq}. Heun's polynomials in particular arise as particular solutions which are regular at three singular points \cite{Hortacsu:2011rr}.
One also demands that the correct solutions are capable of reproducing the eigenfunctions for the Laplacian operator in the standard Schwarzschild background upon taking the limit $w_a\to0$.
\\
In particular, the two classes of solutions respecting all the aforementioned characteristics can be written as follows for $\ell\geq1$
\begin{equation}
\label{eq: U_V}
\begin{split}
    U_\ell(w_a;w)&=\frac{(2M_\alpha)^{\ell+1}}{\ell(\ell+1)}\frac{\Gamma(\ell+2)^2}{\Gamma(2\ell+1)}(1-w_a)^\ell ~ H_l\left(\frac{1}{1-w_a},\frac{\ell(\ell+1)}{w_a-1},-(1+\ell),\ell,1,1;\frac{w-1}{w_a-1}\right)~~,
    \\
    V_\ell(w_a;w)&=2M_\alpha(2\ell+1)
    ~U_\ell(w_a;w)\int_w^\infty\frac{t^2}{(t-1)(t-w_a)U^2_\ell(w_a;t)}dt~~.
\end{split}
\end{equation}
We stress that this are new and original results obtained in this work, representing radial harmonics for vacuum electromagnetic fields around black holes with a Reissner-Nordstr\"{o}m metric structure.
One can check explicitly that for $w_a\to0$ the standard Schwarzschild vacuum fields in terms of hypergeometric functions are recovered \cite{Gralla:2015vta}.
More generally, the function $U_\ell(w_a,w)$ admits a truncated power series representation, akin to
\begin{equation}
    U_\ell(w_a;w)= \frac{(2M_\alpha)^{\ell+1}}{\ell(\ell+1)}\frac{\Gamma(\ell+2)^2}{\Gamma(2\ell+1)}(1-w_a)^\ell \sum_{n=0}^{\ell+1} a^{(\ell)}_n(w_a)\left(\frac{w-1}{w_a-1}\right)^n~~,
\end{equation}
whose coefficients $a^{(\ell)}_n(w_a)$ obey the two-terms recurrence relation
\begin{equation}
\label{eq: recurrence_a}
    \begin{split}
        a^{(\ell)}_{-1}=0~,~~
        a^{(\ell)}_1=1~,~~
        a^{(\ell)}_n=\frac{Q_n a^{(\ell)}_{n-1}+R_na^{(\ell)}_{n-2}}{P_n}~,
    \end{split}
\end{equation}
where 
\begin{equation}
\begin{split}
    P_n&=-\frac{n^2}{w_a-1}~~,~~R_n=(3-n+\ell)(-2+n+\ell) ~~,
    \\
    Q_n&=\frac{\ell(\ell+1)+(n-1)\big[3(1-w_a)-n(2-w_a)\big]}{w_a-1}~~.
\end{split}
\end{equation}
From the series representation above one can explicitly compute the polynomials. For later convenience we list below the expressions of the polynomials for both $U_\ell$ and $V_\ell$ corresponding to $\ell=1,2$:
\begin{align}
\nonumber
        U_1(w_a;w)&=(2M_\alpha)^2(w^2-w_a)
\\\nonumber
        U_2(w_a;w)&=(2M_\alpha)^3\left[w^3-\frac{3}{4}w^2(1+w_a)+\frac{1}{4}w_a(1+w_a)\right]
\\
        V_1(w_a;w)&=-\frac{3}{(2M_\alpha)(1-w_a)^2}\left[w+\frac{1}{2}(w_a+1)+\frac{U_1(w_a;w)}{(2M_\alpha)^2(1-w_a)}\log\left(\frac{w-1}{w-w_a}\right)\right]
\\\nonumber
        V_2(w_a;w)&=-\frac{80}{(2M_\alpha)^2(1-w_a)^4}\left[w^2-\frac{1}{4}w(w_a+1)-\frac{1}{24}[1+w_a(10+w_a)]\right.
\\\nonumber
        &\left.\hspace{40mm}+\frac{U_2(w_a;w)}{(2M_\alpha)^3(1-w_a)}\log\left(\frac{w-1}{w-w_a}\right)\right]
\end{align}
\\

We notice that at the event horizon, $w\to1$, the functions $U_\ell$ are properly defined, whereas $V_\ell$ manifests a logarithmic divergence
\begin{equation}
\begin{split}
    U_\ell(w_a,1)&= \frac{(2M_\alpha)^{\ell+1}}{\ell(\ell+1)}\frac{\Gamma(\ell+2)^2}{\Gamma(2\ell+1)}(1-w_a)^\ell
    ~~,
    \\
    ~~
    V_\ell(w_a,1)&\sim\frac{U_\ell(w_a,1)}{[2M_\alpha(1-w_a)]^{2\ell+1}}\log\left(\frac{w-1}{w-w_a}\right)~~.
\end{split}
\end{equation}
Viceversa, for $w\to\infty$ the asymptotic values are
\begin{equation}
    \label{eq: Asymptotic_UV}
    U_\ell(w_a,\infty)\sim (2M_\alpha w)^{\ell+1}~~,~~V_\ell(w_a,\infty)\sim (2M_\alpha w)^{-\ell}~~.
\end{equation}
Hence the vacuum radial functions $U_\ell$ are convergent at the horizon and divergent at infinity, whereas the opposite holds for the funtions $V_\ell$.

\subsection{Asymptotically monopolar static field}
\label{sec: static 4.3}

As an example of vacuum magnetic field solutions around static MOG background \eqref{eq: SchwMOG}, we can consider configurations with an isotropic distribution of magnetic field lines at infinity. A monopolar magnetosphere is specified by the following asymptotic boundary condition \cite{Grignani:2019dqc}
\begin{equation}
\label{eq: BCmonopole}
    \lim_{r\to\infty}\Psi=\Psi^{\infty}(\theta)~~.
\end{equation}
Among the exact solutions we derived, there exists a monopole magnetic field solution given by 
\begin{equation}\label{eq:homogenous}
\Psi(r,\theta)\equiv \psi_0(\theta)=1-  \cos\theta~~,
\end{equation}
where the integration constant has been chosen in order to respect the normalisation condition of the flux in Eq.~\eqref{eq: BC0}.\\

A monopolar magnetosphere around a static MOG black hole is thus indistinguishable from one surrounding a Schwarzschild black hole in GR.
However, as we detail in the next section, this is no longer true for a monopole magnetosphere around spinning black holes, with the static solution constructed here that will serve as the starting point to extend the BZ perturbative procedure to the Kerr-MOG case.

\subsection{Asymptotically vertical static field}

Before moving to consider the case of a stationary MOG background we present here another interesting topology for a vacuum field.
Following the definition given in \cite{Grignani:2019dqc}, a vertical topology \cite{Pan:2017npg} is defined by the condition that $\Psi$ remains finite for $r\to\infty$ and the product $r\cdot \theta$ is kept fixed. This can be more easily visualised in cartesian coordinates, $r=\sqrt{x^2+z^2}$ and $\cos\theta=z/\sqrt{x^2+z^2}$, with the asymptotic boundary condition that now reads
\begin{equation}
\label{eq: BCvertical}
    \lim_{z\to\infty}\Psi=\Psi^{\infty}(x)~~.
\end{equation}
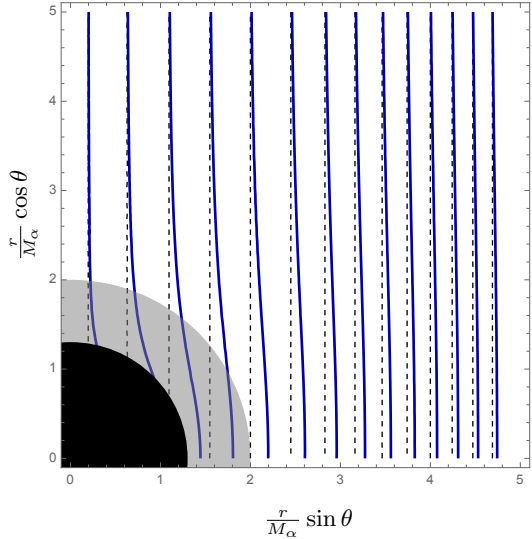
\begin{figure}
    \centering
    \includegraphics[scale=0.5]{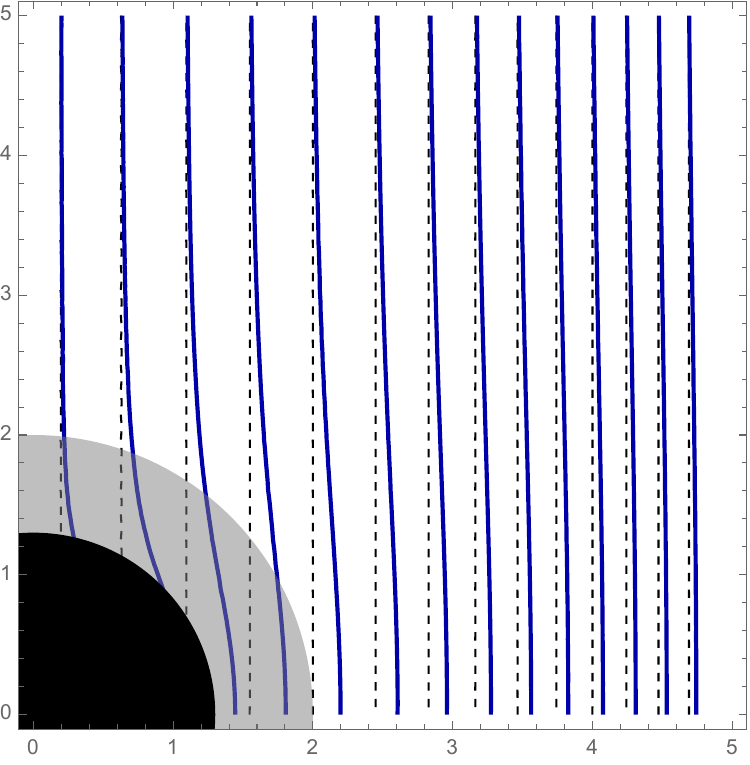}
    \\
    \begin{picture}(0,0)
        \put(-10,0){\footnotesize{$\frac{r}{M_\alpha}\sin\theta$}}
        \put(-105,100){\rotatebox{90}{\footnotesize{$\frac{r}{M_\alpha}\cos\theta$}}}
    \end{picture}
    \caption{Asymptotically vertical magnetostatic field lines (blue) around a static MOG black hole. For comparison in the same plot the case of a vertically uniform field (dashed black) around a Schwarzschild black hole in GR -- whose horizon is depicted in gray -- is reported. The plot assumes $\alpha=10$ to magnify the distorsion of the magnetic field lines in the vicinity of the non-spinning MOG black hole. }
    \label{fig:Vertical}
\end{figure}
By considering vacuum solutions of the type $\Psi(r,\theta)=U_{\ell}(r)\Theta_{\ell}(\theta)$ which are regular at the event horizon, and by converting to cartesian coordinates, it is possible to observe that for large values of $z$ one has $\Psi(x,z)\sim  x^2 z^{\ell-1}$. Thus, the only vacuum solution which is regular both at the event horizon and at infinity and that is consistent with the asymptotically vertical boundary condition \eqref{eq: BCvertical}, corresponds to $\ell=1$ and explicitly reads
\begin{equation}
    \Psi(r,\theta)=\frac{r^2(1+\alpha)-\alpha M_\alpha^2}{2M_\alpha^2(1+\sqrt{1+\alpha})}\sin^2\theta~~,
\end{equation}
where an integration constant was fixed by means of the normalisation condition in Eq.~\eqref{eq: BC0}. We notice that for $\alpha\to0$ one recovers a vertical magnetic field configuration in the Schwarzschild background with profile given by $\Psi=r^2/(2M)^2\sin^2\theta$ \cite{Pan:2017npg}.\\

An interesting result of this paper is that, in contrast with the GR case, vacuum magnetic fields surrounding static MOG black holes, and characterised by a vertical asymptotic profile, are not \emph{uniformly vertical}. In Fig.~\ref{fig:Vertical} we present an illustrative comparison between the asymptotically vertical magnetic field for a Schwarzschild and a non-spinning MOG black hole with the same ADM mass. The derivation obtained here shows that, interestingly, deviations from GR not only affect the dynamics at large scales, but can also be observed in strong-gravity magnetostatic configurations. This might suggest that also the geometrical factor $\kappa$ in the BZ formula \eqref{eq: BZ_0} for the energy extracted can receive modification by the presence of the MOG parameter $\alpha$, as opposed to the case of a monopole magnetosphere where only the factor $f(\Omega_H)$ will change.

\section{Force-free Kerr-MOG magnetospheres}
\label{sec: BZMOG}

In this section we generalise the BZ approach \cite{Blandford:1977ds,Armas:2020mio,Camilloni:2022kmx} for the construction of a split-monopole magnetosphere in the Kerr-MOG background in the regime of slow rotation, namely for small values of the black hole spin $\epsilon$. At each order in perturbation theory we provide explicit expressions for the magnetospheric field variables, and 
we comment about the regularity of the solution across all the critical surfaces that characterise the magnetospheric problem.

\subsection{Leading order solution}

Since in the limit $\epsilon \to 0$ the Kerr-MOG metric \eqref{eq:KerrMOG} reduces to the static MOG metric \eqref{eq: SchwMOG}, a perturbative solution of the Grad-Shafranov equation \eqref{eq: GS_eq} for small values of $\epsilon$ can be constructed order by order in an $\epsilon$ expansion,
by starting from a vacuum field in the static MOG background, $F_{\mu\nu}\sim \mathcal{O}(\epsilon^0)$, with an associated current which is assumed to scale as $j^\mu\sim\mathcal{O}(\epsilon)$. At the leading order in the expansion for small $\epsilon$ then the field and the vector current automatically satisfy the force-free constraint, $F_{\mu\nu}j^\nu\sim\mathcal{O}(\epsilon)$. These assumptions are satisfied as long as the force-free field variables scale as
\begin{equation}
    \label{eq: scaling}
    \Psi\sim\psi_0+\mathcal{O}(\epsilon)~~,~~\Omega\sim\mathcal{O}(\epsilon)~~,~~I\sim\mathcal{O}(\epsilon)
\end{equation}
In this section we only consider the split-monopole configuration
\begin{equation}
    \psi_0=1-\cos\theta~~,
\end{equation}
which has been derived in the previous section after specifying the boundary conditions \eqref{eq: BC0} and \eqref{eq: BCmonopole}.\\

With the assumptions \eqref{eq: scaling}, the condition $g^{\mu\nu}\eta_\mu\eta_\nu=0$ that determine the light surfaces reveal the following scaling in the spin parameter $\epsilon$
\begin{equation}
\label{scalingILSOLS}
        \frac{r_{\rm ILS}}{M_\alpha}\sim1+\frac{1}{\sqrt{1+\alpha}}+\mathcal{O}(\epsilon)~~,~~
        \frac{r_{\rm OLS}}{M_\alpha}\sim \frac{1}{\epsilon}+\mathcal{O}(\epsilon^0)~~.
\end{equation}
In other words, at the leading order, the ILS coincide with the event horizon, whereas the OLS is located at infinity.
The non-perturbative scaling of the OLS is at the core of the Matched Asymptotic Expansion scheme that is needed to consistently construct higher-order solutions in the BZ perturbation theory \cite{Armas:2020mio,Camilloni:2022kmx}. In the case of a split-monopole field in Kerr the necessity of the MAE scheme becomes evident at the fourth order in perturbation theory \cite{Armas:2020mio,Camilloni:2022kmx}. We expect the same to be true in Kerr-MOG. However, in order to disentangle possible modified gravity effects from specific configurations of the magnetosphere \cite{Dong:2021yss} it turns out to be sufficient to truncate the series at the third order in the perturbative expansion, so that in the present work we do not need to deal with the MAE scheme.
\\

All the quantities will be normalised to the black hole gravitational mass $M_\alpha$, which is independent from the expansion parameter $\epsilon$ and will facilitate the comparison with results for Kerr black holes with same ADM mass as the Kerr-MOG black hole under consideration.

\subsection{First order in the small spin regime}

The ansatz for the field variables expansion at the first subleading order is
\begin{equation}
    \begin{split}
        &\Psi=\psi_0(\theta)+\epsilon~ \psi_1(r,\theta)+\mathcal{O}(\epsilon^2)~~,
        \\
        &I(\Psi)=\frac{\epsilon}{2M_\alpha}~i_1(\psi_0)+\mathcal{O}(\epsilon^2)~~,
        \\
        &\Omega(\Psi)=\frac{\epsilon}{2M_\alpha}~\omega_1(\psi_0)+\mathcal{O}(\epsilon^2)~~,
    \end{split}
\end{equation}
where a factor proportional to the ADM mass of the black hole is included so as to have dimensionless perturbative coefficients, and the integrability conditions \eqref{eq: intcond} have been used to constrain the dependence of $i_1$ and $\omega_1$.

\subsubsection{Magnetic flux}
The function $\psi_1$ obeys the source-less stream equation $\mathcal{L}\psi_1=0$, and the unique solution which is regular at both the horizon and at infinity is the trivial solution $\psi_1=0$.

\subsubsection{Poloidal current and angular velocity of the field lines}
By means of the Znajek condition \eqref{eq: ZC} at the event horizon one obtains
\begin{equation}
\label{eq: ZCH_i1}
    i_1(\theta)=\left[\frac{2(1+\alpha)}{(1+\sqrt{1+\alpha})^2}-\omega_1(\theta)\right]\Theta_1(\theta)~~.
\end{equation}
\\
The asymptotic Znajek condition \eqref{eq: ZC} instead gives 
\begin{equation}
\label{eq: ZCinf_i1}
    i_1(\theta)=\omega_1(\theta)\Theta_1(\theta)~~.
\end{equation}
By comparing the two conditions above one obtains explicit expressions for the current and the angular velocity at this order in perturbation theory
\begin{equation}
    \label{eq: i1omega1}
    i_1(\theta)=\frac{1+\alpha}{(1+\sqrt{1+\alpha})^2}~\Theta_1(\theta)~~,~~\omega_1=\frac{1+\alpha}{(1+\sqrt{1+\alpha})^2}~~.
\end{equation}
Notice that in the limit $\alpha\to0$, these two quantities reproduce the results known in the literature for the Kerr spacetime, namely $i_1=1/4~\Theta_1(\theta)$ and $\omega_1=1/4$.
\\

The expressions in Eq.s~\eqref{eq: ZCH_i1} and \eqref{eq: ZCinf_i1} can also be derived by demanding regularity of the stream equation respectively at the ILS and at OLS, by means of \eqref{eq: stream_LS} \cite{Camilloni:2022kmx, Camilloni2022}. As already explained, for $\epsilon=0$, the ILS and the OLS are located at the event horizon and at infinity respectively. Turning on the spin parameter $\epsilon$, it is possible to compute perturbatively the new location of the ILS and OLS. For the first correction we find 
\begin{equation}
    \begin{split}
        \frac{r_{\rm ILS}}{M_\alpha}&=1+\frac{1}{\sqrt{1+\alpha}}-\epsilon^2\frac{\sqrt{1+\alpha}}{2}\left(1-\frac{\sin^2\theta}{4}\right)+\mathcal{O}(\epsilon^3)
        \\
        \frac{r_{\rm OLS}}{M_\alpha}&=\frac{1}{\epsilon}\left(1+\frac{1}{\sqrt{1+\alpha}}\right)^2\frac{2}{ \sin\theta}+\mathcal{O}(\epsilon^0)
    \end{split}
\end{equation}
By evaluating the reduced stream equation \eqref{eq: stream_LS} at the new location for the ILS and OLS, one gets an equation for $i_1$ and $\omega_1$ which is precisely solved by the expressions given in  Eq.~\eqref{eq: i1omega1}. 
Thus, all the quantities defined at this order in perturbation theory are regular at all the critical surfaces.

\subsection{Second order in the small spin regime}

Moving to the next perturbative order, the expansion now reads
\begin{equation}
    \begin{split}
        &\Psi=\psi_0(\theta)+\epsilon^2~ \psi_2(r,\theta)+\mathcal{O}(\epsilon^3)~~,
        \\
        &I(\Psi)=\frac{\epsilon}{2M_\alpha}\left[i_1(\psi_0)+\epsilon~ i_2(\psi_0)\right]+\mathcal{O}(\epsilon^3)~~,
        \\
        &\Omega(\Psi)=\frac{\epsilon}{2M_\alpha}\left[\omega_1(\psi_0)+\epsilon~ \omega_2(\psi_0)\right]+\mathcal{O}(\epsilon^3)~~,
    \end{split}
\end{equation}
where the dependence on $i_2$ and $\omega_2$ can be directly inferred from the integrability conditions \eqref{eq: intcond}.

\subsubsection{Magnetic flux}

Expanding the stream equation \eqref{eq: GS_eq} up to $\mathcal{O}(\epsilon^2)$ and using the known expressions for $\psi_0$, $\omega_1$ and $i_1$, one obtains the following equation for the function $\psi_2(r,\theta)$
\begin{equation}
    \mathcal{L}\psi_2(r,\theta)=-4\frac{M_\alpha^3}{r^4\bar r_+^2}\left(r-\frac{\bar r_+\bar r_-}{2M_\alpha}\right)\left(\frac{r+\bar r_+}{r-\bar r_-}\right)\frac{\Theta_2(\theta)}{\sin\theta}~~
\end{equation}
where we recall that $\bar r_\pm=M_\alpha\left(1\pm\sqrt{(1+\alpha)^{-1}}\right)$ label the outer/inner horizon positions in the case of a static MOG black hole (see Sec. \ref{sec: static}).\\

By assuming the solution to be separable, $\psi_2(r,\theta)=R_2(r)\Theta_2(\theta)$, one can project this equation on $\Theta_2(\theta)$ by using the orthogonality conditions which characterise the angular harmonics $\Theta_\ell(\theta)$. This produces a non-homogeneous differential equation for the radial part which can be conveniently written in terms of the dimensionless radial coordinate $w=r/\bar r_+$ and $w_a=\bar r_-/\bar r_+$, introduced in Eq.~\eqref{eq: w_def}, as
\begin{equation}
    \begin{split}
    R''_2(w)-&\left(\frac{2}{w}-\frac{1}{w-1}-\frac{1}{w-w_a}\right)R_2'(w)-\frac{6}{(w-1)(w-w_a)}R_2(w)
    \\
    &\hspace{20mm}=-\frac{(w+1)(w_a+1)^3}{2(w-1)w^2(w-w_a)^2}\left(w-\frac{w_a}{w_a+1}\right)~~.
    \end{split}
\end{equation}
This equation is the same type of Heun's equation that we studied in the vacuum static case (See Eq.~\eqref{eq: Heuneq}), with $\ell=2$ and with an additional non-homogeneous term. It is possible to determine analytically a particular solution to this equation which is regular at both the event horizon ($w=1$) and at infinity after properly adding the homogeneous solutions $U_2(w_a;w)$ and $V_2(w_a;w)$. The explicit solution reads 
\begin{equation}
\label{eq: R2}
\begin{split}
    R_2(w)=&-\frac{1}{2w_a}\frac{(1+w_a)^2}{(1-w_a)^2}\Bigg\{2 w^2(1+3w_a)-\frac{w(1+w_a)(3+w_a)}{2}
    -\frac{w_a[17+w_a(20-w_a)]}{9}
    \\
    &-\frac{6w[4w-(1+w_a)]-[1+w_a(10+w_a)]}{6}\left[\log (w)-(1+w_a)\log\left(\frac{w-w_a}{1-w_a}\right)\right]+
    \\
    &-\frac{4w^3-(1+w_a)(3w^2-w_a)}{1-w_a}\left[\mbox{Li}_2\left(\frac{w_a}{w}\right)-\mbox{Li}_2\left(\frac{1}{w}\right)+\log( w  )\log \left(\frac{w-1}{w-w_a}\right)+\right.
    \\
    &\left.+(1+w_a)\left(\frac{\pi^2}{6}+\frac{1}{2}\log^2\left(\frac{w-w_a}{1-w_a}\right)-\frac{(1-w_a)}{2w_a}\log\left(1-\frac{w_a}{w}\right)+\mbox{Li}_2\left(\frac{1-w}{1-w_a}\right)\right)\right]\Bigg\}
\end{split}
\end{equation}
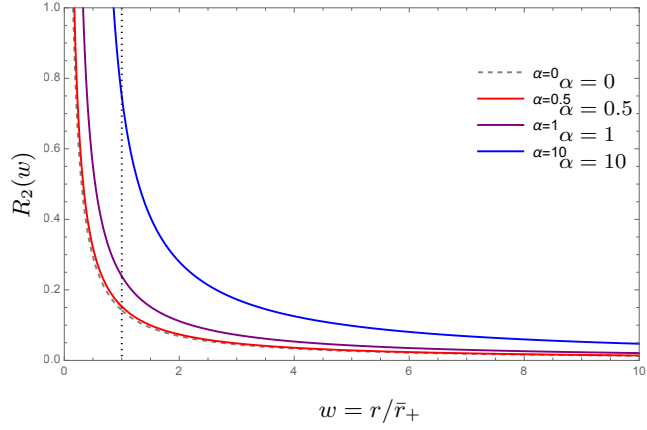
\begin{figure}
    \centering
    \includegraphics[scale=0.45]{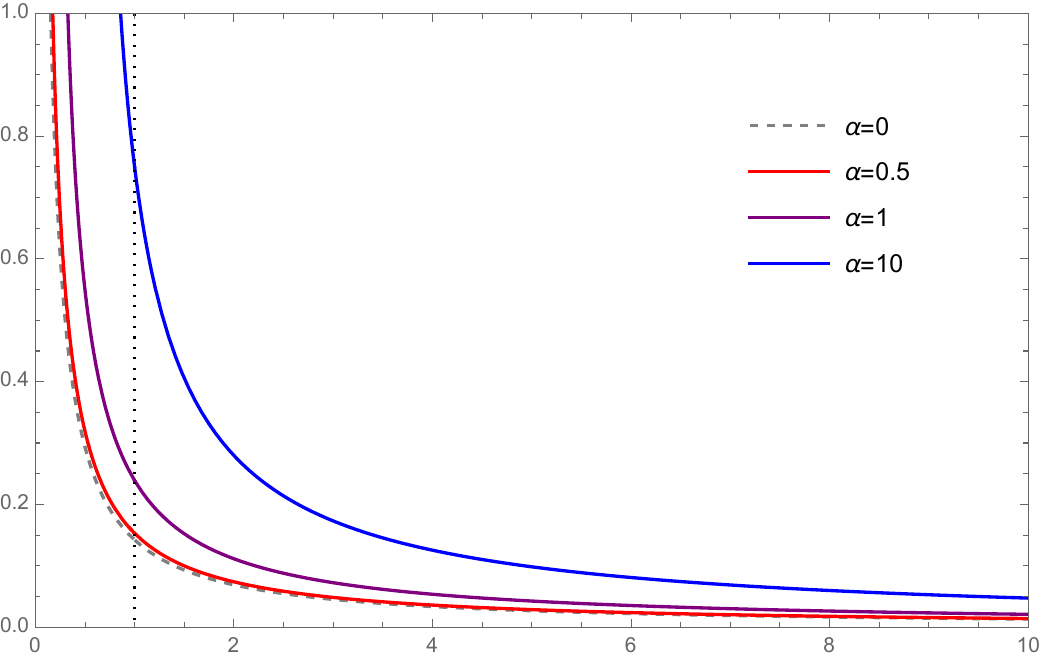}
        \\
    \begin{picture}(0,0)
        \put(-10,0){\footnotesize{$w=r/\bar{r}_+$}}
        \put(-125,75){\rotatebox{90}{\footnotesize{$R_2(w)$}}}
        \put(80,124){\scriptsize{$\alpha=0$}}
        \put(80,114){\scriptsize{$\alpha=0.5$}}
        \put(80,104){\scriptsize{$\alpha=1$}}
        \put(80,94){\scriptsize{$\alpha=10$}}
    \end{picture}
    \caption{Plot for the function $R_2(w)$, obtained by varying the MOG parameter $\alpha$. The dashed line represents the function $R_2(w)$ in the GR case, Eq.~\eqref{eq: R2GR}. The dotted vertical line marks the position of the static event horizon at $w=1$, namely $r=\bar r_+$. }
    \label{fig: R2}
\end{figure}
We illustrate in Fig.~\ref{fig: R2} the behaviour of the function $R_2(w)$ for some representative values of the deformation parameter $\alpha$. 
Notice that by taking the limit $w_a\to0$, and after using the inversion formula for the dilogarithms, one has
\begin{equation}
    \label{eq: R2GR}
\begin{split}
    R_2(w)=\frac{11}{72}&+\frac{1}{6w}+w(1-2w)+\frac{1+6w-24w^2}{12}\log(w)
    \\
    &+\frac{w^2(4w-3)}{2}\left[\mbox{Li}_2\left(\frac{1}{w}\right)+\log w \log\left(1-\frac{1}{w}\right)\right]~~,
\end{split}
\end{equation}
which is precisely the result known in the Kerr spacetime.

The function $R_2(w)$ is smooth in the asymptotic region
\begin{equation}
 R^\infty_2=\frac{(1+w_a)^3}{8w}+\mathcal{O}\left(\frac{1}{w^2}\right)~~,
\end{equation}
which trivially reproduces the result known in the Kerr spacetime $R^\infty_2=\frac{M}{4r}$ in the limit $\omega_a\to0$ \cite{Blandford:1977ds,Camilloni:2022kmx}. This means that the solution at $\mathcal{O}(\epsilon^2)$ extends all the way up to the asymptotic region, smoothly crossing the OLS.

By making use of the inversion relation one can also infer  the behaviour at the static horizon $w=1$, \emph{i.e.} $r=\bar r_+=1+\sqrt{(1+\alpha)^{-1}}$. This simply reads
\begin{equation}
\label{eq: R2H}
    R^H_2=\frac{(1+w_a)^2}{2(1-w_a)w_a}\left[-\frac{1}{2}+\frac{w_a(3\pi^2-47+2w_a)}{18}+\mbox{Li}_2\left(w_a\right)-\frac{(1-w_a^2)}{2w_a}\log(1-w_a)\right]~~.
\end{equation}
Remarkably, the limit $w_a\to0$ is finite and reproduces the value obtained expanding up to second order in the spin parameter in the case of a monopole solution for the Kerr  magnetosphere, $R^H_2=\frac{6\pi^2-49}{72}$ \cite{Camilloni:2022kmx}. We also recall that explicitly one has $w_a=\frac{1+\alpha-\sqrt{1+\alpha}}{1+\alpha+\sqrt{1+\alpha}}$ for the Kerr-MOG spacetime.

\subsubsection{Poloidal current and angular velocity of the field lines}
The Znajek condition at the horizon, expanded at the order $\epsilon^2$, gives
\begin{equation}
\label{eq: ZCH_i2}
    i_2(\theta)=-\omega_2(\theta)\Theta_1(\theta)~~.
\end{equation}
Similarly, the Znajek condition at infinity at the order $\epsilon^2$ gives
\begin{equation}
    i_2(\theta)=\omega_2(\theta)\Theta_1(\theta)~~.
\end{equation}
By comparing the two relations above, one immediately has for consistency that 
\begin{equation}
    \label{eq: i2omega2}
    i_2(\theta)=\omega_2(\theta)=0~~.
\end{equation}

The corrections to the positions of the light surfaces can be obtained by demanding that $r_{\rm erg}(\theta)\geq r_{\rm ILS}(\theta)\geq r_+$ which, together with Eq.~\eqref{eq: i2omega2}, leads to the following explicit expressions for the light surface positions at order $\epsilon^2$ 
\begin{equation}
    \begin{split}
        \frac{r_{\rm ILS}}{M_\alpha}&=1+\frac{1}{\sqrt{1+\alpha}}-\epsilon^2\frac{\sqrt{1+\alpha}}{2}\left(1-\frac{\Theta_1(\theta)}{4}\right)+\mathcal{O}(\epsilon^4)~~,
        \\
        \frac{r_{\rm OLS}}{M_   \alpha}&=\frac{1}{\epsilon}\left(1+\frac{1}{\sqrt{1+\alpha}}\right)^2\frac{2}{ \sin\theta}-1+\mathcal{O}(\epsilon)~~.
    \end{split}
\end{equation}
%

\subsection{Third order in the small spin regime}

By including terms of order $\epsilon^3$, the expansion of the force-free field variables reads
\begin{equation}
    \begin{split}
        &\Psi=\psi_0(\theta)+\epsilon^2~ \psi_2(r,\theta)+\epsilon^3~ \psi_3(r,\theta)+\mathcal{O}(\epsilon^4)~~,
        \\
        &I(\Psi)=\frac{\epsilon}{2M_\alpha}\left[i_1(\psi_0)+\epsilon^2\left(\frac{\partial_\theta i_1}{\partial_\theta \psi_0}\psi_2 (r,\theta)+ i_3(\psi_0)\right)\right]+\mathcal{O}(\epsilon^4)~~,
        \\
        &\Omega(\Psi)=\frac{\epsilon}{2M_\alpha}\left[\omega_1(\psi_0)+\epsilon^2~ \omega_3(\psi_0)\right]+\mathcal{O}(\epsilon^4)~~.
    \end{split}
\end{equation}
Again, the integrability conditions $I=I(\Psi)$ and $\Omega=\Omega(\Psi)$, Eq.~\eqref{eq: intcond},  dictate the structure of the expansion at the third order in perturbation theory.

\subsubsection{Magnetic flux}

Similarly to what we saw at the first perturbative order, the function $\psi_3$ obeys a sourceless stream equation $\mathcal{L}\psi_3=0$. This means that the trivial solution $\psi_3=0$ is the only one consistent with the boundary conditions of the split-monopole configuration.

\subsubsection{Poloidal current and angular velocity of the field lines}

Expanding the Znajek condition at the horizon \eqref{eq: ZC} up to order $\epsilon^3$, one obtains a relation between $i_3$ and $\omega_3$ of the type
\begin{equation}
\label{eq: ZCH_i3}
\begin{split}
    i_3(\theta)=&-\omega_3(\theta)\Theta_1(\theta)
    \\
    &-\frac{1}{5}\frac{1+\alpha}{(1+\sqrt{1+\alpha})^2}\left(R^H_2-\frac{1+\alpha}{(1+\sqrt{1+\alpha})^2}\right)\Theta_3(\theta)
    \\
    &-\frac{2(1+\alpha)}{(1+\sqrt{1+\alpha})^2}\left[\frac{2}{5}R_2^H-\frac{1+\alpha}{(1+\sqrt{1+\alpha})^2}\left(\frac{7}{5}+\frac{\alpha}{1+\sqrt{1+\alpha}}\right)\right]\Theta_1(\theta)~~.
  \end{split}
\end{equation}
The asymptotic Znajek condition at the third perturbative order, instead, simply produces
\begin{equation}
\label{eq: ZCinf_i3}
    i_3(\theta)=\omega_3(\theta)\Theta_1(\theta)~~.
\end{equation}
By comparing the two equations above one obtains the following expression for $i_3(\theta)$
\begin{equation}
\label{eq: i3}
    \begin{split}
    i_3(\theta)=&-\frac{1+\alpha}{(1+\sqrt{1+\alpha})^2}\left[\frac{2}{5}R_2^H-\frac{1+\alpha}{(1+\sqrt{1+\alpha})^2}\left(\frac{7}{5}+\frac{\alpha}{(1+\sqrt{1+\alpha})}\right)\right]\Theta_1(\theta)
    \\
    &-\frac{1}{10}\frac{1+\alpha}{(1+\sqrt{1+\alpha})^2}\left(R^H_2-\frac{1+\alpha}{(1+\sqrt{1+\alpha})^2}\right)\Theta_3(\theta)~~,
  \end{split}
\end{equation}
while for $\omega_3(\theta)$ we have
\begin{equation}
\label{eq: omega3}
    \begin{split}
    \omega_3(\theta)=&\frac{(1+\alpha)^2}{(1+\sqrt{1+\alpha})^4}\left(1+\frac{\alpha}{1+\sqrt{1+\alpha}}\right)
    \\
    &+\frac{1}{2}\frac{(1+\alpha)^2}{(1+\sqrt{1+\alpha})^4}\left[1-\frac{R_2^H}{1+\alpha}\left(4+2\alpha-\frac{\alpha^2}{(1+\sqrt{1+\alpha})^2}\right)\right]\Theta_1(\theta)~~.
  \end{split}
\end{equation}
In the limit $\alpha\to0$ Eq.s \eqref{eq: i3} and \eqref{eq: omega3} correctly reproduce the results for the Kerr magnetosphere, namely \cite{Camilloni:2022kmx}
\begin{equation}
\begin{split}
    i_3(\theta)&=\frac{7-8R^H_2}{80}\Theta_1(\theta)+\frac{1-4R_2^H}{160}\Theta_3(\theta)~~,
    \\
    \omega_3(\theta)&=\frac{1}{16}+\frac{1-4R_2^H}{32}\Theta_1(\theta)~~,
\end{split}
\end{equation}
where in the GR case $R_2^H=\frac{6\pi^2-49}{72}$.\\

The position of ILS now read 
\begin{equation}
\label{eq: rILS3}
\begin{split}
    \frac{r_{\rm ILS}}{M_\alpha}=&1+\frac{1}{\sqrt{1+\alpha}}-\epsilon^2\frac{\sqrt{1+\alpha}}{2}\left(1-\frac{\Theta_1(\theta)}{4}\right)
    \\
    &-\epsilon^4\frac{(1+\alpha)^{3/2}}{8}\left[1-\left(1+\frac{8\sqrt{1+\alpha}}{1+\sqrt{1+\alpha}}+\frac{16R_2^H}{1+\alpha}\right)\frac{\Theta_1(\theta)}{20}\right.
    \\
    &\left.\hspace{15mm}+\left(9-\frac{8\sqrt{1+\alpha}}{1+\sqrt{1+\alpha}}-\frac{16R_2^H}{1+\alpha}\right)\frac{\Theta_3(\theta)}{80}\right]+\mathcal{O}(\epsilon^5)~~,
\end{split}
\end{equation}
whereas the OLS location is
\begin{equation}
\label{eq: rOLS3}
\begin{split}
    \frac{r_{\rm OLS}}{M_\alpha}&=\frac{1}{\epsilon}\left(1+\frac{1}{\sqrt{1+\alpha}}\right)^2\frac{2}{ \sin\theta}-1
    \\
    &-\frac{\epsilon}{\sin\theta}\left[2\sqrt{1+\alpha}+\left(\frac{(3+2\alpha)(1-\sqrt{1+\alpha})}{2\alpha^2}\right. \right.
    \\
    &\left.\left.\hspace{10mm}-\frac{(1+\sqrt{1+\alpha})^2}{1+\alpha}R_2^H+\frac{3(2\alpha+1)}{4\alpha}\right)\Theta_1(\theta)\right]+\mathcal{O}(\epsilon^2)~~.
\end{split}
\end{equation}
Notably, the limit $\alpha\to0$ yields a finite result that correctly reproduces the light surfaces locations for the Kerr magnetosphere \cite{Camilloni:2022kmx} at the third perturbative order in the spin parameter. 
It is possible to verify by direct substitutions that $i_3$ and $\omega_3$ as defined in Eq.s~\eqref{eq: i3} and \eqref{eq: omega3}, satisfy the stream equation when evaluated at the locations \eqref{eq: rILS3} and \eqref{eq: rOLS3}. The solution obtained, hence, is regular at the horizon and at infinity, implying regularity at the ILS and OLS as well.

\subsection{Spinning split-monopole in Kerr-MOG and bunching of field lines}

We conclude this section by presenting plots of the analytic solution we perturbatively derived for a split-monopole magnetosphere in a slowly rotating Kerr-MOG background up to third order in the spin parameter $\epsilon$.

\begin{figure*}
    \centering
    \includegraphics[scale=0.55]{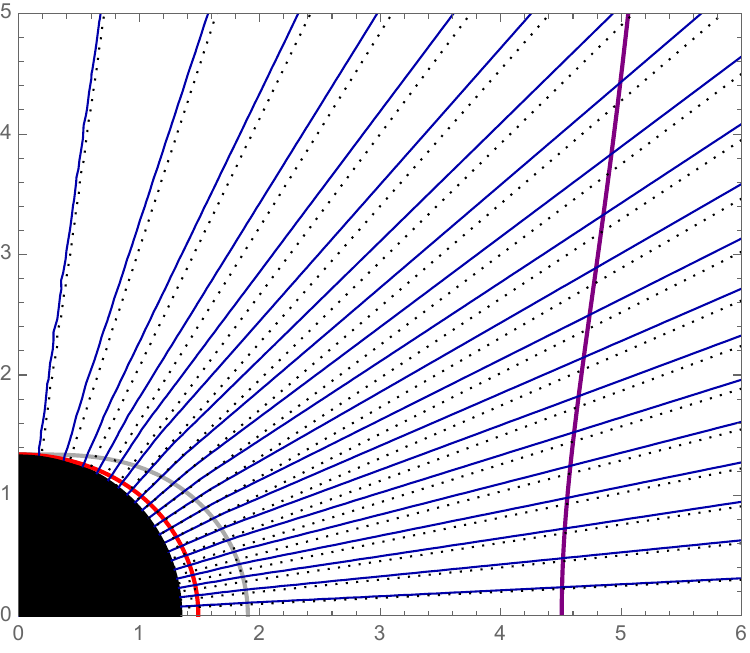}
    \hspace{1mm}\includegraphics[scale=0.65]{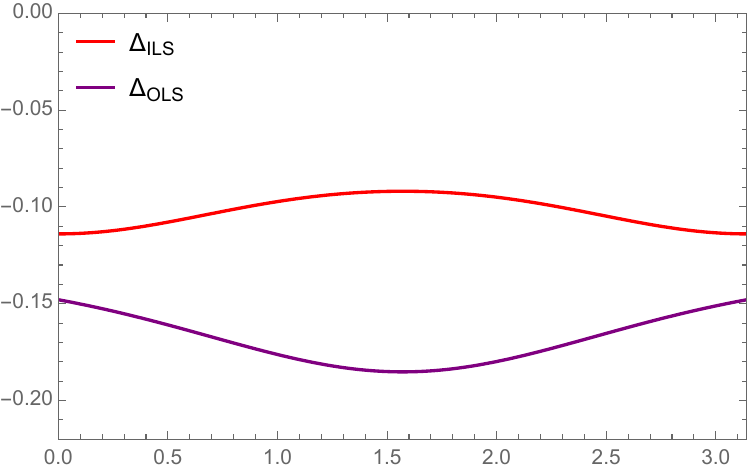}
    \\
    \begin{picture}(0,0)
        \put(25,145){$\footnotesize{\Delta_{\rm ILS}}$}
        \put(25,133){$\footnotesize{\Delta_{\rm OLS}}$}
        \put(108,5){$\footnotesize{\theta}$}
        \put(-145,5){\footnotesize{$\frac{r}{M_\alpha}\sin\theta$}}
        \put(-235,85){\rotatebox{90}{\footnotesize{$\frac{r}{M_\alpha}\cos\theta$}}}
    \end{picture}
    \caption{\textit{Left Panel:} the magnetic flux $\Psi$ for a monopolar configuration around a Kerr-MOG black hole, plotted together with the magnetospheric surfaces of interest: red for the ILS, purple for the OLS, gray for the ergosphere and black for the event horizon. The magnetic field lines correspond to curves of constant $\Psi$, depicted in blue for $\epsilon=0.9$ and as dotted lines in the static case $\epsilon=0$. The plot has been obtained by fixing $\alpha=0.23$.
    \textit{Right Panel:} plot for the fractional deviation $\Delta_{\rm X}=(r_X(\alpha)-r_X(0))/r_X(0)$ of the ILS and OLS positions in the case $\epsilon=0.9$ and $\alpha=0.23$, with respect to the GR case $\alpha=0$. The negative values of the fractional deviations for both the ILS and the OLS indicate that the critical surfaces are closer to the black hole in the Kerr-MOG case. }
    \label{fig: Monopole}
\end{figure*}
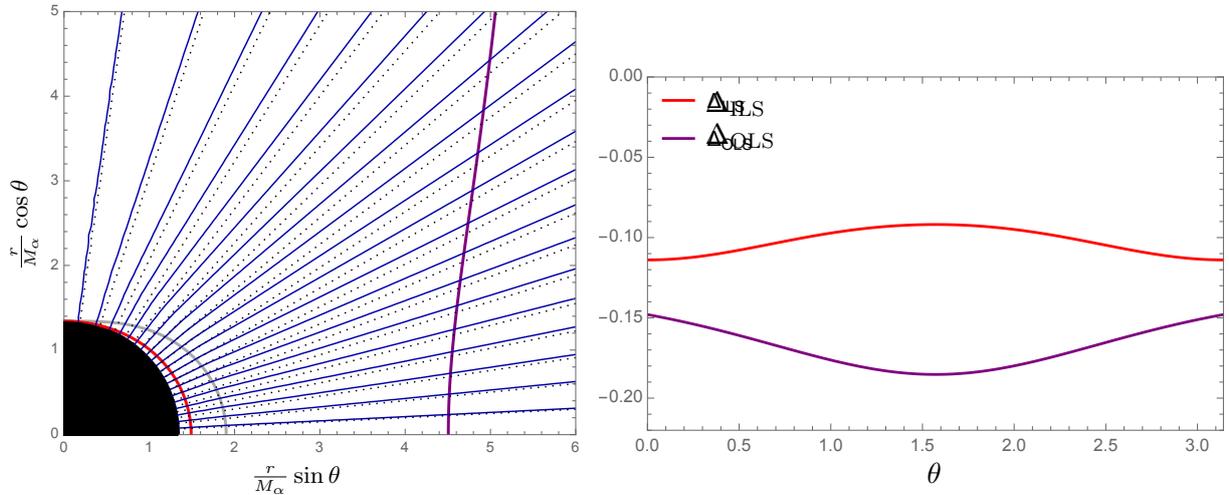

It appears evident from the left panel of Fig.~\ref{fig: Monopole} that the magnetic field lines are smooth at both the ILS and OLS, whose respective locations are given analytically in Eq.s~\eqref{eq: rILS3} and \eqref{eq: rOLS3}. It is possible to observe that, because of the enhanced gravitational attraction of a Kerr-MOG black hole, the light surfaces lie closer to the event horizon compared to the case of the magnetosphere for a GR Kerr black hole of the same ADM mass. This is displayed in the right panel of Fig.~\ref{fig: Monopole}, where the fractional deviations of the positions of the light surfaces with respect to the GR case is plotted as a function of the polar angle.

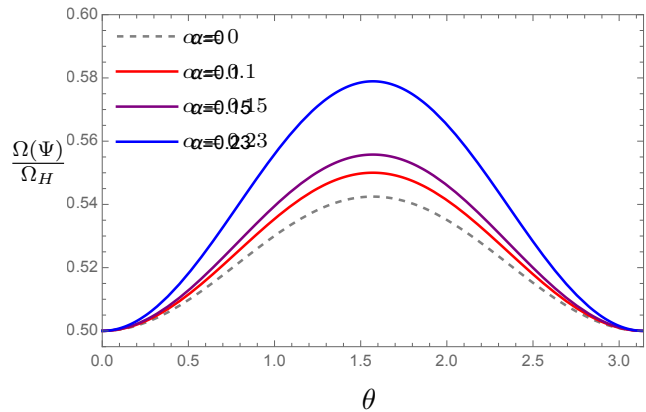
\begin{figure}
    \centering
    \includegraphics[scale=0.6]{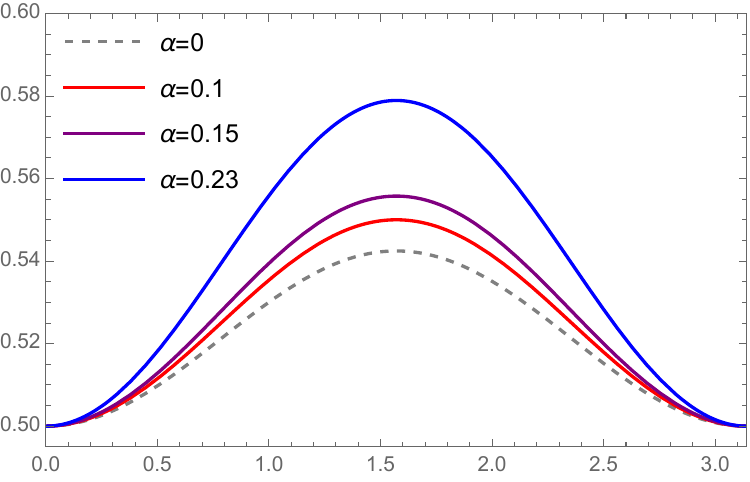}
        \\
    \begin{picture}(0,0)
        \put(2,0){$\theta$}
        \put(-130,90){$\frac{\Omega(\Psi)}{\Omega_H}$}
        \put(-65,138){\scriptsize{$\alpha=0$}}
        \put(-65,125){\scriptsize{$\alpha=0.1$}}        
        \put(-65,112){\scriptsize{$\alpha=0.15$}}
        \put(-65,99){\scriptsize{$\alpha=0.23$}}
    \end{picture}
    \caption{Angular distribution for the velocity of the magnetic field lines $\Omega(\Psi)$ in the monopolar case, for four different values of the MOG parameter, $\alpha=0,0.1,0.15,0.23$, and for $\epsilon=0.9$. Notice that the value $\alpha\approx 0.23$ approximately corresponds to the maximal value of the MOG parameter when the spin of the Kerr-MOG black hole is fixed to $\epsilon=0.9$. Viceversa $\alpha=0$ corresponds to a Kerr black hole. The plot has been obtained by converting the expansion in the spin parameter $\epsilon$ into an expansion in the black hole angular velocity $\Omega_H$, $\Omega=\omega_1 \Omega_H+\omega_3 \Omega_H^3+\mathcal{O}(\Omega_H^4)$. }
    \label{fig: Omega}
\end{figure}

In Fig.~\ref{fig: Omega} the ratio between the angular velocity of the magnetic field lines $\Omega(\Psi)$ and the angular velocity of the black hole $\Omega_H$ is plotted with respect to the polar angle $\theta$. The comparison with the GR case ($\alpha=0$) makes it clear that a monopolar magnetosphere around a Kerr-MOG black hole spins faster when compared to the Kerr case.
\\

Numerical \cite{2010ApJ...711...50T} as well as analytical \cite{Gralla:2015uta} studies of magnetospheres in Kerr spacetime lead to the observation that the magnetic field lines tend to bunch up towards the rotational axis $\theta=0$ when the black hole is in the high-spinning regime.
With the magnetospheric solution derived here, we are now able to investigate whether this \emph{bunching of field lines} also occur around a Kerr-MOG black hole.

To this aim one can compute the contravariant component of the radial magnetic field which, according to Eq.~\eqref{eq: F_FFE}, is related to the magnetic flux through $B^r=\partial_\theta \Psi/\sqrt{-g}$. By converting the expansion in the spin parameter $\epsilon$ into an expansion for a dimensionless angular velocity at the horizon $\omega_H=M_\alpha \Omega_H $, the second-order accurate expression for $B^r$ reads
\begin{equation}
    B^r=\frac{1+\left(1+\frac{1}{\sqrt{1+\alpha}}\right)^4\omega_H^2R_2(r)(2-3\Theta_1(\theta))}{\frac{r^2}{M_\alpha^2}+\left(\frac{2+\alpha+2\sqrt{1+\alpha-\alpha^2\omega_H^2}}{(1+\alpha)(1+4\omega_H^2)}\right)^2\omega_H^2\cos^2\theta}~~,
\end{equation}
which directly reduces to the expression known in the Kerr metric when $\alpha\to0$ \cite{2010ApJ...711...50T}.

\begin{figure}
    \centering
    \includegraphics[scale=0.55]{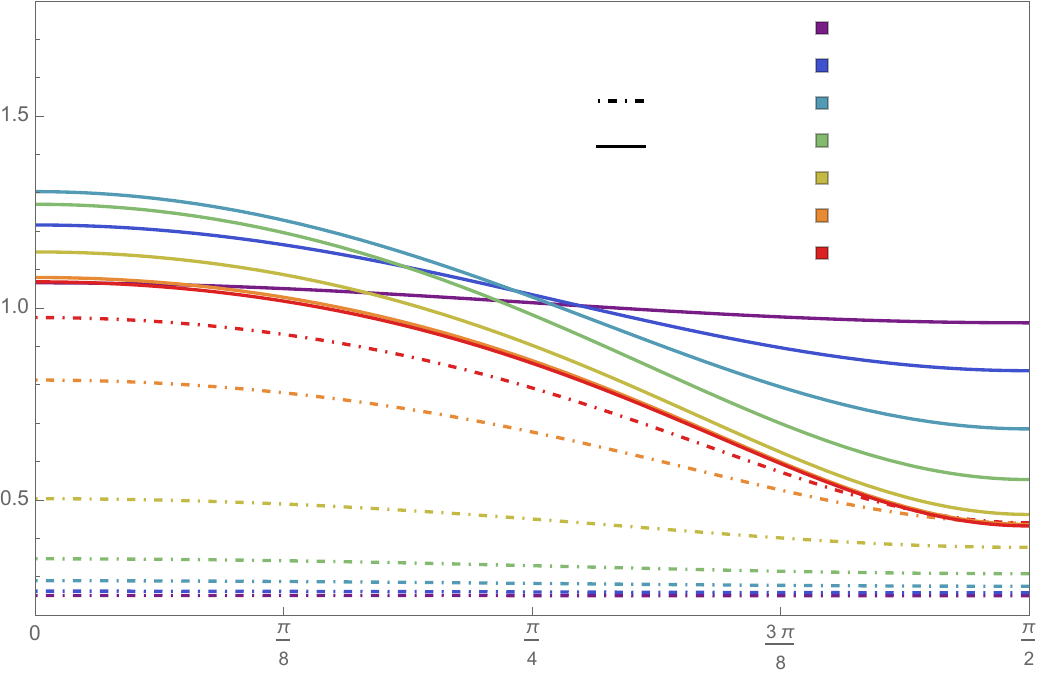}
        \\
    \begin{picture}(0,0)
        \put(-5,3){$\theta_H$}
        \put(-155,115){$B^r_H$}
        \put(10,163){{\scriptsize{$\alpha=0$}}}
        \put(10,151){{\scriptsize{$\alpha^*=\epsilon^{-2}-1$}}}
        \put(84,182){{\scriptsize{$\epsilon=0.1$}}}
        \put(84,173){{\scriptsize{$\epsilon=0.3$}}}
        \put(84,163){{\scriptsize{$\epsilon=0.5$}}}
        \put(84,153){{\scriptsize{$\epsilon=0.7$}}}
        \put(84,143){{\scriptsize{$\epsilon=0.9$}}}
        \put(84,133){{\scriptsize{$\epsilon=0.99$}}}
        \put(84,123){{\scriptsize{$\epsilon=0.999$}}}
    \end{picture}
    \caption{The colours represent different spin parameters, whereas dot-dashed and solid lines distinguish between the GR case and the maximal value of $\alpha$ consistent with \eqref{eq:PhysicalBound}. The increasing of the value for $B^r$ at $\theta=0$ is considered as a signature of the bunching of field lines. }
    \label{fig: Bunch}
\end{figure}
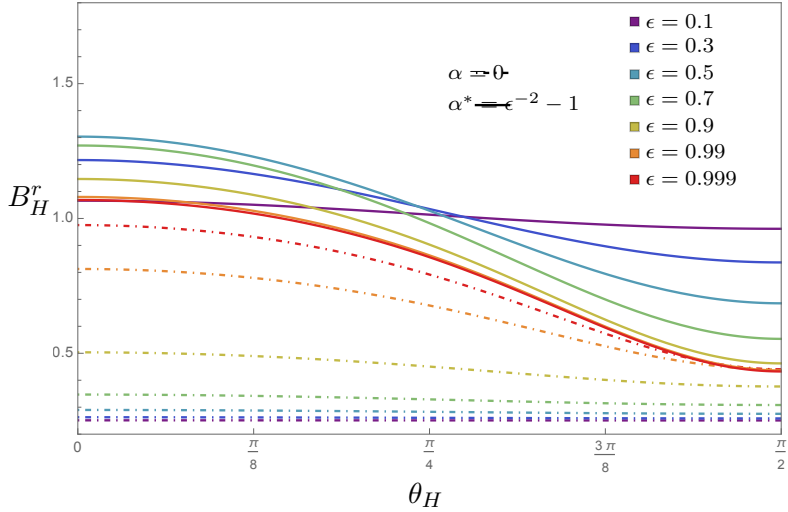

In Fig.~\ref{fig: Bunch} we present a plot for $B^r$ evaluated at the horizon $r=r_+$, obtained by varying the MOG deformation parameter $\alpha$ and the angular velocity of the Kerr-MOG black hole $\omega_H$. In particular, we present a comparison between the GR case, $\alpha=0$, and the case in which $\alpha$ takes its maximum value allowed for a given black hole spin, \emph{i.e.} $\alpha^*=\epsilon^{-2}-1$, (see Eq.~\eqref{eq:PhysicalBound}). The former case is depicted in the plot with dot-dashed curves, for which $\omega_H=\epsilon/(2+2\sqrt{1-\epsilon^2})$, whereas for $\alpha=\alpha^*$ solid curves are adopted and one has $\omega_H^*=\epsilon/(1+\epsilon^2)$. All possible intermediate values for $\alpha$, thus, lie within dot-dashed and solid curves of the same colour in Fig.~\ref{fig: Bunch}.

The figure makes immediate to observe that increasing the angular velocity of the Kerr-MOG black hole and the MOG deformation parameter leads to two competitive effects. For fixed black hole spin, indeed, $B^r_H$ at $\theta=0$ increases as the deformation parameter $\alpha$ approaches the maximum value. In other words, MOG deviations from GR contribute in a positive manner to the bunching of the field lines towards the rotational axis of the black hole. As one spins the black hole up, the maximum value for $B^r_H$ at $\theta=0$ is reached for $\omega\approx0.42$, namely $\epsilon\approx 0.54$. After this, the value of $B^r_H$ decreases until it reaches $\omega_H\to1/2$ (corresponding to $\alpha=\alpha^*$ and $\epsilon\to 1$), where the window of allowed values for $\alpha$ narrows down to $\alpha\to0$. In this limit, clearly, the value of $B^r_H$ attains the corresponding value for an extreme Kerr black hole.
%

\section{Blandford--Znajek mechanism in Kerr-MOG}
\label{sec: BZPower}

We assume that the jet is black-hole powered, and that most of its energy is extracted by means of the BZ mechanism \cite{Blandford:1977ds}. The perturbative solution we derived for the magnetosphere can be therefore exploited to compute the rate of energy and angular momentum extracted from the Kerr-MOG black hole.\\

In particular, the power and the angular momentum per unit of time extracted at the horizon can be computed by means of the following integrals \cite{Gralla:2014yja,Camilloni:2022kmx}
\begin{equation}
\label{enexp}
\begin{split}
    \dot{E}(r_+)&=2\pi\int_0^\pi \Omega(r_+,\theta) I(r_+,\theta)\partial_\theta\Psi(r_+,\theta) d \theta~~,
    \\
    \dot{L}(r_+)&=2\pi\int_0^\pi I(r_+,\theta)\partial_\theta\Psi(r_+,\theta) d \theta~~.
\end{split}
\end{equation}
By making use of the expansions of the field variables in the spin parameter $\epsilon$ one can write the expressions above as a series expansion in $\epsilon$ as follows
\begin{equation}
\begin{split}
    \dot{E}(r_+)&=\epsilon^2~\dot{E}_+(\psi_0,i_1,\omega_1)+\epsilon^4~\dot{E}_+(\psi_{0,2},i_{1,3},\omega_{1,3})+\mathcal{O}(\epsilon^6)~~,
    \\
    \dot{L}(r_+)&=\epsilon~\dot{L}_+(\psi_0,i_1)+\epsilon^3~\dot{L}_+(\psi_{0,2},i_{1,3})+\mathcal{O}(\epsilon^4)~~.
\end{split}
\end{equation}
where inside the parenthesis we have explicitly written the dependence on the coefficients of the expansion for $\Psi$, $I$ and $\Omega$ that contribute to the integrals in Eq.~\eqref{enexp} at the specific order in the $\epsilon$ expansion.
\\
By means of Eq.~\eqref{eq:Omega_H} it is possible to trade the expansion in the dimensionless spin parameter $\epsilon$ for an expansion in the angular velocity of the black hole $\Omega_H$, and arrive at the more familiar expressions for the BZ rate of energy and angular momentum extraction at the horizon of a Kerr-MOG black hole
\begin{equation}
\label{eq: BZpower}
    \dot E=\frac{2\pi}{3}\Omega_H^2 f^E_\alpha(\Omega_H)~~,~~\dot L=\frac{4\pi}{3}\Omega_H f^L_\alpha(\Omega_H)~~ .
\end{equation}
As mentioned in the introduction, the prefactors are related to the characteristic geometrical quantity $\kappa=\frac{2\pi}{3}\cdot\frac{1}{4\pi^2}\approx 0.053$ of a monopolar magnetosphere (that, as already inferred in sec.~\ref{sec: static 4.3}, remains unaltered with respect to the GR case), whereas the deviation functions $f^{E,L}_\alpha(\Omega_H)$, up to $\mathcal{O}(\Omega_H^2)$, read
\begin{equation}
    \label{eq: f}
    f^{E,L}_\alpha(\Omega_H)=1+\frac{2}{5}M_\alpha^2\Omega_H^2 C^{E,L}(\alpha)+\mathcal{O}(\Omega_H^4)
\end{equation}
and the explicit expressions for $C^{E,L}(\alpha)$ are given by
\begin{equation}
\label{eq: C}
\begin{split}
    C^E(\alpha)&=2\frac{(1+\sqrt{1+\alpha})^2}{1+\alpha}\left(1-\frac{(1+\sqrt{1+\alpha})^2}{1+\alpha}R_2^H(\alpha)\right)~~,
    \\
    C^L(\alpha)&=\frac{1}{2}C^E(\alpha)~~.
\end{split}
\end{equation}
We recall that $R_2^H(\alpha)$ is given in Eq.~\eqref{eq: R2H} as a function of $\alpha$, which in the GR case reduces to $R_2^H=\frac{6\pi^2-49}{72}$. In other words, the expression for the deviation function $f^E_\alpha(\Omega_H)$ above is a highly non-linear function of the deformation parameter $\alpha$ which, in the limit $\alpha\to0$, correctly reproduces $f^E_0\approx1+1.3835 (M \Omega_H)^2+\mathcal{O}(\Omega_H^4)$, as previously obtained for Kerr black holes \cite{2010ApJ...711...50T,Camilloni:2022kmx}. An analogous argument holds for the function $f^L_\alpha(\Omega_H)$.
It is interesting to notice that, at this order in perturbation theory, the deviation functions for the energy and angular momentum extraction rates share the same dependence on the deformation parameter $\alpha$, and differ only by means of numerical factors.

It is immediate to recognise that the leading term in Eq.~\eqref{eq: BZpower} is precisely the same that one would have obtained for the BZ mechanism derived in \cite{Blandford:1977ds} for the magnetosphere surrounding a Kerr black hole. Such a degeneracy was first noted in \cite{Dong:2021yss}\footnote{In \cite{Konoplya:2021qll} it was shown that the quadratic scaling in $\Omega_H$ at the leading order of $\dot E_+$ is characteristic of every axially symmetric, asymptotically flat, spinning black hole background.} to affect the leading order term of the power and angular momentum extracted for the BZ mechanism in alternative theories of gravity. Effects of modified gravity and of specific configurations for a black hole magnetospheres can therefore be isolated and disentangled only by considering subleading contributions in the BZ mechanism, that enters at order $\Omega_H^2$ in the factor $f^E_\alpha(\Omega_H)$.
\\
\begin{figure}
    \centering
    \includegraphics[scale=0.5]{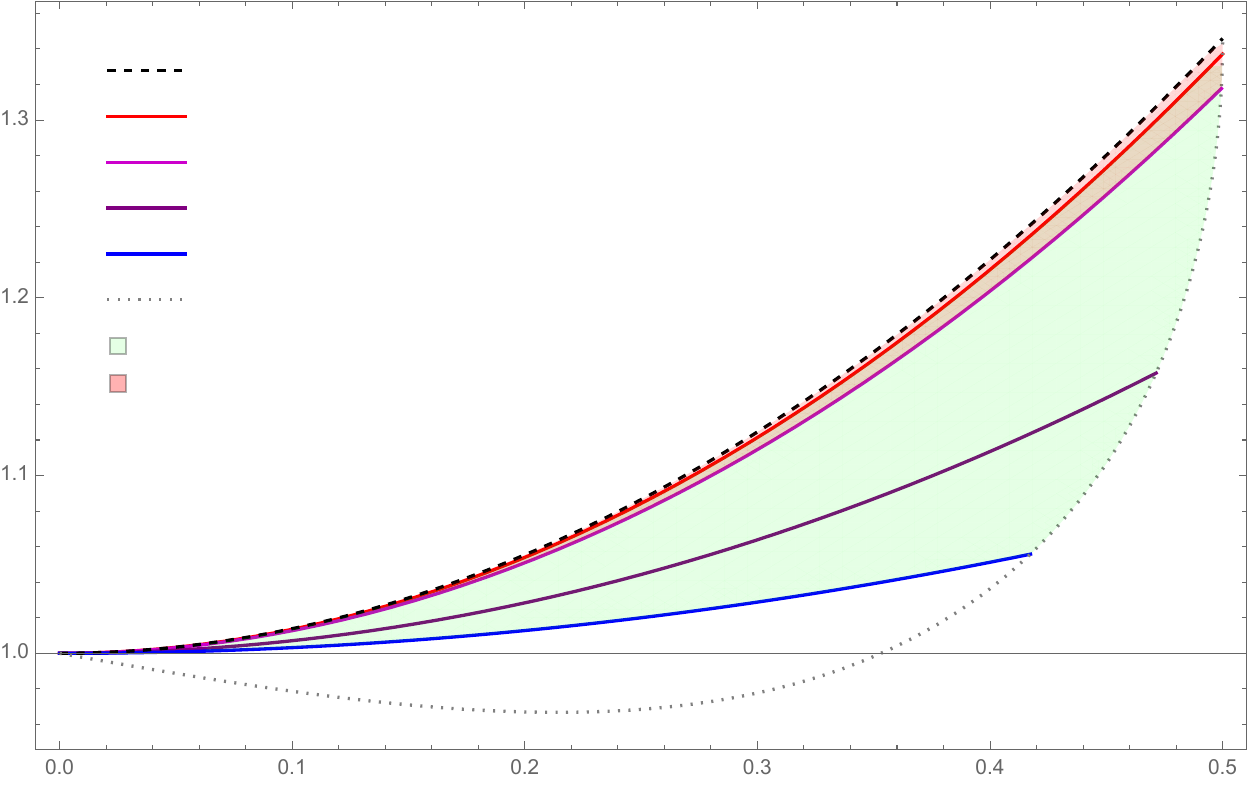}
    \\
    \begin{picture}(0,0)
        \put(-25,0){{$\omega_H=M_\alpha \Omega_H$}}
        \put(-165,110){\rotatebox{90}{{$f^E_\alpha$}}}
        \put(-100,183){{\scriptsize{$(\alpha,\omega^*)=(0,0.5)$}}}
        \put(-100,172){{\scriptsize{$(\alpha,\omega^*)=(0.03,0.4999)$}}}
        \put(-100,161){{\scriptsize{$(\alpha,\omega^*)=(0.1,0.4994)$}}}
        \put(-100,150){{\scriptsize{$(\alpha,\omega^*)=(1,0.4714)$}}}
        \put(-100,139){{\scriptsize{$(\alpha,\omega^*)=(2.45,0.4174)$}}}
        \put(-100,128){{\scriptsize{$\alpha^*=1/\epsilon^2-1$}}}
        \put(-115,118){{\scriptsize{$\alpha\in [0.03,2.47]$ SMBH}}}
        \put(-115,108){{\scriptsize{$\alpha<0.1$ Stellar BH}}}
    \end{picture}
    \caption{Plot for the function $f_\alpha$, given in Eq.~\eqref{eq: f} and characterising the BZ mechanism in the MOG background, as a function of the dimensionless horizon angular velocity $\omega_H=M_\alpha \Omega_H $. The curves are obtained by varying the MOG parameter $\alpha$ for specific reference values. In particular, the dashed black line corresponds to the GR case, $\alpha=0$. Because of the physical bound \eqref{eq:PhysicalBound}, the curves in the MOG case truncate at the maximum spin value $\epsilon^*=1/\sqrt{1+\alpha}$, $\omega^*=\frac{\sqrt{1+\alpha}}{2+\alpha}$, displayed in the legend together with the corresponding value of $\alpha$. The dotted line stands for the curve of all the maximum values for the power extracted from a Kerr-MOG black hole at extremality.
    The region between the dashed and the dotted curve is the region accessible for the MOG case, according to the physical constraints \eqref{eq:PhysicalBound}. More specifically, the green and red areas respectively correspond to estimated range of values for $\alpha$ in the case of supermassive \cite{Brownstein:2005dr,Moffat:2007yg}
    and stellar mass
    \cite{LopezArmengol:2016irf}
    MOG black holes.}
    \label{fig: fOmega}
\end{figure}
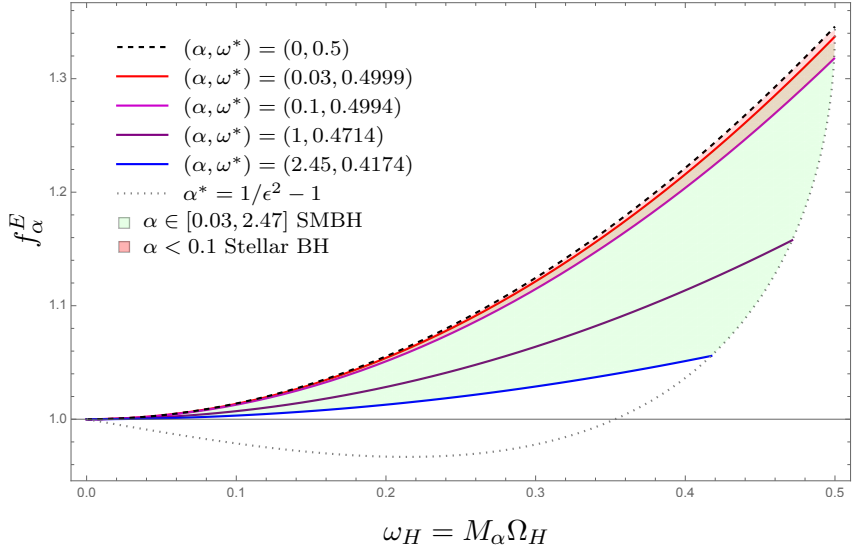

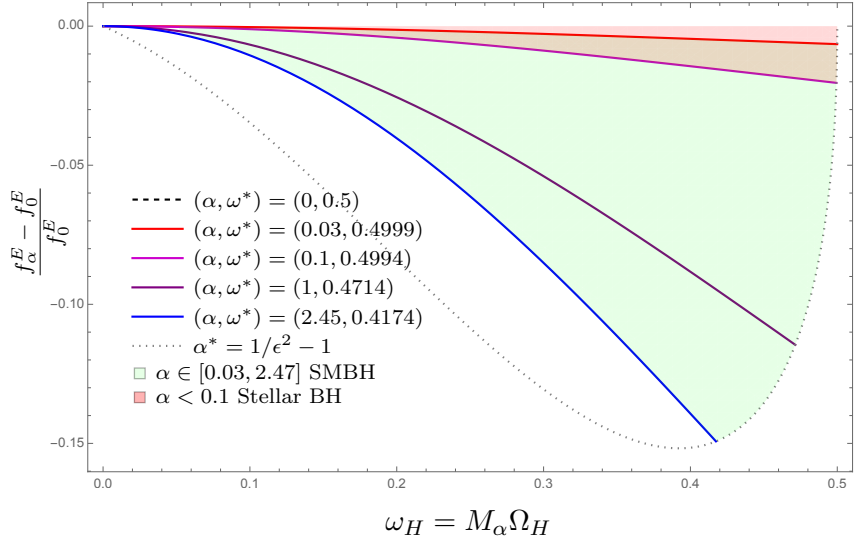
\begin{figure}
    \centering
    \includegraphics[scale=0.5]{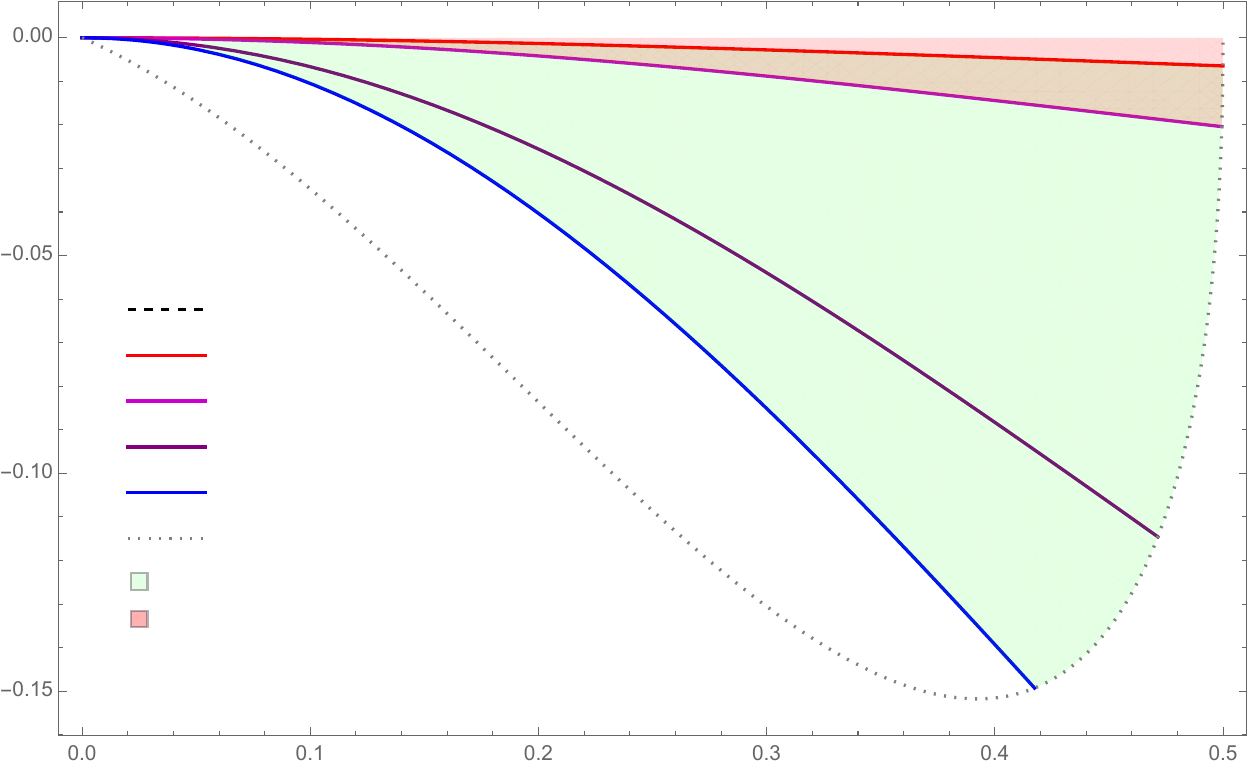}
    \\
    \begin{picture}(0,0)
        \put(-25,0){{$\omega_H=M_\alpha \Omega_H$}}
        \put(-165,95){\rotatebox{90}{{\scriptsize{$\dfrac{f^E_\alpha-f^E_0}{f^E_0}$}}}}
        \put(-97,122){{\scriptsize{$(\alpha,\omega^*)=(0,0.5)$}}}
        \put(-97,111){{\scriptsize{$(\alpha,\omega^*)=(0.03,0.4999)$}}}
        \put(-97,100){{\scriptsize{$(\alpha,\omega^*)=(0.1,0.4994)$}}}
        \put(-97,89){{\scriptsize{$(\alpha,\omega^*)=(1,0.4714)$}}}
        \put(-97,78){{\scriptsize{$(\alpha,\omega^*)=(2.45,0.4174)$}}}
        \put(-97,67){{\scriptsize{$\alpha^*=1/\epsilon^2-1$}}}
        \put(-111,57){{\scriptsize{$\alpha\in [0.03,2.47]$ SMBH}}}
        \put(-111,48){{\scriptsize{$\alpha<0.1$ Stellar BH}}}
    \end{picture}
    \caption{ Fractional deviation of the function $f^E_\alpha$ from the GR expression $f^E_0$ truncated at $\mathcal{O}( \Omega_H^2)$. The deviations from the GR factor $f^E_0$ are of order $\leq -5 \%$ in the case of stellar mass black hole, whereas they can attain $\leq -14 \%$ in the case of supermasssive black holes.}
    \label{fig: PowerOmega}
\end{figure}

The plot in Fig.~\ref{fig: fOmega} displays the function $f^E_\alpha(\Omega_H)$ according to Eq.~\eqref{eq: f} as a function of the black hole angular velocity $\Omega_H$ and for specific values of the MOG parameter $\alpha$. 
In particular, the values in the parameter space have been chosen to be consistent with previous estimates for the deformation parameter that can be found in the literature \cite{Perez:2020ndx}. For stellar mass Kerr-MOG black holes \cite{LopezArmengol:2016irf} derived an upper limit $\alpha<0.1$ (light red zone in Fig.~\ref{fig: fOmega}). For supermassive Kerr-MOG black holes the deformation parameter lies in the range $\alpha\in[0.03,2.47]$ (light green zone in Fig.~\ref{fig: fOmega}), with the upper limit obtained in \cite{Brownstein:2005dr} to reproduce the rotational curves of white dwarf galaxies, and the lower limit derived in \cite{Moffat:2007yg} to study globular cluster velocity dispersion.
\\
As it is clear from the picture if we consider a Kerr and a Kerr-MOG black hole of the same ADM mass and angular velocity $\Omega_H$, the power extracted at the horizon is reduced in the Kerr-MOG case compared to the result for a Kerr black hole.
\\
In Fig.~\ref{fig: PowerOmega} we plot the relative deviation of $f^E_\alpha$ with respect to its GR limit $f^E_0$. Interestingly, the relative deviations become more relevant in the region of the parameter space which corresponds to MOG black holes in the supermassive regime, which precisely constitute the primary candidates for EHT observations.\\
Our results in Eq.s~\eqref{eq: BZpower}, \eqref{eq: f} and \eqref{eq: C} (illustrated in Fig.s \ref{fig: fOmega} and \ref{fig: PowerOmega}) show that by combining high-precision estimates of the jet power with independent measurements of the black hole spin or angular frequency, it is possible to probe the metric of astrophysical black holes and possibly put constraints on deformation parameters \cite{Pei:2016kka}.
In the present work, specifically, we focused on the Kerr-MOG scenario, and obtained a non-degenerate expression for the BZ power emitted at order $\mathcal{O}(\Omega_H^4)$ without making assumptions on the magnitude of the deformation parameter $\alpha$.
This constitutes an advancement with respect to the current literature about the BZ mechanism in alternative theories of gravity, which either truncated the expression for the power emitted at the leading order \cite{Banerjee:2020ubc,Konoplya:2021qll} or exploited a double expansion in both small spin and small deformation parameters to derive next-to-leading order results \cite{Dong:2021yss,Peng:2023}.
%

\section{Concluding remarks}
\label{sec: conclusion}

The main purpose of this work is to study the BZ mechanism around Kerr-MOG black holes \cite{Moffat:2014aja}. In order to accomplish this goal several intermediate results have been achieved. 

More specifically, in Sec.~\ref{sec: static} we analytically classified all vacuum static magnetic field configurations around non-spinning black holes in MOG in terms of angular harmonics and radial Heun's polynomials \cite{Ronveaux:1995:HDE}. It is important to stress that these results are solely based on the singularity structure of the Laplacian operator in metrics akin to the Reissner-Nordstr\"{o}m metric. We therefore envision that the solutions here derived can also be useful in studying magnetic field configurations around electrically charged black holes in the test field limit \cite{Komissarov:2021vks}.
We explicitly showed that, while the solution for a static monopolar vacuum field in MOG is indistinguishable from its GR counterpart, the case with vertical asymptotic topology is qualitatively different when compared to the case of a Schwarzschild black hole in the strong-gravity region. We expect that this difference can reflect in the geometrical factor $\kappa$ present in the expression for the energy extracted in the BZ mechanism, though further investigations are needed in this direction.

In Sec.~\ref{sec: BZMOG} we considered the BZ perturbative approach \cite{Blandford:1977ds,Armas:2020mio,Camilloni:2022kmx} in order to construct the first analytical model for a spinning monopolar magnetosphere in a Kerr-MOG background, up to the third order in the perturbative expansion. At each order in perturbation theory we proved the smoothness of the solution across all the critical surfaces characterising the magnetospheric problem, and we studied how the presence of a MOG deformation parameter contributes in a positive manner to the bunching of the field lines towards the rotational axis of the black hole. 

Having an analytical description of black hole magnetospheres is important and interesting in its own right. First of all  since our understanding of the energy extraction mechanism is incomplete, and only through an analytical model one can attain a deeper understanding. Moreover, the analytical model are complementary to the numerical simulations. In the analytical model one can directly obtain the dependence on the key parameters such as the black hole angular velocity, whereas the simulations can only cover one set of parameters at a time. 
In addition, the analytic solution derived here can be beneficial to adapt GRMHD codes which exploits force-free approximation and stationarity, and to perform numerical simulations of black hole magnetospheres in the Kerr-MOG background.

Finally, in Sec.~\ref{sec: BZPower} the explicit expression for the power extracted at the horizon of a Kerr-MOG black hole in the BZ mechanism was computed. We showed that its expression, $\dot E=\frac{2\pi}{3}\Omega_H^2 f_\alpha(\Omega_H)$, is formally similar to the one obtained in the Kerr black hole background. In fact, in the case of monopolar magnetospheres, only the function $f_\alpha(\Omega_H)$, that accounts for deviations from a quadratic dependence on the angular velocity $\Omega_H$, allows to distinguish the MOG case from the standard GR case.
As an important result, we showed with an explicit example that the expression for $f(\Omega_H)$ depends on the specific theory of gravity on which the BZ mechanism is set to operate.
For the Kerr-MOG case, we derived $f_\alpha(\Omega_H)$ up to orders $\mathcal{O}(\Omega_H^4)$, as explicitly given in Eq.~\eqref{eq: f}, and the subregion of the parameter space within which the function can vary is depicted in Fig.~\ref{fig: fOmega}. In Fig.~\ref{fig: PowerOmega} we also showed that the fractional deviation of $f_\alpha(\Omega_H)$ from the expression it takes for the standard Kerr case is more relevant in the range of the MOG parameter $\alpha$ which characterises supermassive black holes. \\

As already emphasized, the analytical approach is relevant in order to obtain a clear understanding of the physics of black hole magnetospheres. Moreover, our model and the expression we obtained for $f_\alpha(\Omega_H)$ can provide analytic support for the construction of novel GRMHD simulations that take into account the MOG deformation parameter, and which can be used to constrain future high-precision horizon-scale observations from the EHT collaboration. In the context of GR it is known that in order to reproduce the numerical data for the power emitted in the BZ mechanism for black holes in the high spin regime further subleading corrections in $\Omega_H$ are needed in the expression of $f(\Omega_H)$ \cite{2010ApJ...711...50T,Camilloni:2022kmx}. Given that GRMHD simulations in the high-spinning regime for black holes are computationally expensive \cite{Talbot:2020zkb}, and that a complete knowledge of the BZ mechanism in modified theories of gravity would require an entire scanning of the parameter space, enhanced by the presence of one or more deformation parameters, analytic models as the one proposed here and higher-orders extensions are expected to provide precious information to overcome these issues. We leave the construction of additional subleading corrections for future works.\\

Finally, while this research focused on the specific case of the MOG scenario, it would be extremely interesting to extend the analysis of the BZ mechanism to theory-agnostic backgrounds, such as the Konoplya-Rezzolla-Zhidenko metric \cite{PhysRevD.93.064015}. A first step in this direction was taken in \cite{Konoplya:2021qll}, even though the analysis was limited to the leading order contribution for the power emitted, which cannot be used to distinguish GR from alternative theories of gravity due to a degeneracy among the spin and deformation parameters, see \cite{Dong:2021yss} and our discussion in Sec. \ref{sec: BZPower}. We plan to investigate theory-agnostic backgrounds in future projects.
The model constructed here should be considered synergetic to future theory-agnostic studies which can use our results to make comparison with the specific MOG case.

\section*{Acknowledgements}

We thank G. Grignani for his contributions in an early stage of this project and R. Oliveri for useful comments.
F.C.~acknowledges support by the ERC Advanced Grant “JETSET: Launching, propagation and emission of relativistic jets from binary mergers and across mass scales” (Grant No. 884631). F.C. and M.O.~acknowledge financial support of the Ministero dell’Istruzione dell’Università e della Ricerca (MUR) through the program “Dipartimenti di Eccellenza 2018-2022” (Grant SUPER-C). M.O.~acknowledge financial support from Fondo Ricerca di Base 2020 (MOSAICO) and 2021 (MEGA) of the University of Perugia. The work of MJR is partially supported through the NSF grant PHY-2012036, RYC-2016-21159, CEX2020- 001007-S and PGC2018-095976-B-C21, funded by MCIN/AEI/10.13039/501100011033. T.H. thanks Perugia University and F.C. and M.O. thank Niels Bohr Institute for hospitality.

\bibliography{[JCAP]Bibliography.bib} 

\end{document}